\documentclass[conference]{IEEEtran}

\newif\ifarxivversion
\arxivversionfalse

\providecommand{\Description}[1]{}

\usepackage{cite}
\usepackage{amsmath,amssymb,amsfonts}
\usepackage{amsthm}
\usepackage{booktabs}
\usepackage{multirow}
\usepackage{enumitem}
\usepackage{placeins}
\usepackage{capt-of}
\usepackage{graphicx}
\usepackage[ruled,vlined,linesnumbered]{algorithm2e}
\usepackage{balance}
\usepackage{hyperref}
\usepackage{cleveref}

\graphicspath{{figures/}}

\newcommand{\ExpTableFont}{\scriptsize}

\newtheoremstyle{fgimdefinition}%
  {.5\baselineskip plus .2\baselineskip minus .2\baselineskip}
  {.5\baselineskip plus .2\baselineskip minus .2\baselineskip}
  {\itshape}
  {0pt}
  {\bfseries}
  {.}
  {.5em}
  {\thmname{#1}\thmnumber{ #2}\thmnote{ \normalfont(#3)}}
\theoremstyle{fgimdefinition}
\newtheorem{definition}{Definition}
\theoremstyle{plain}

\begin{document}

\setlength{\textfloatsep}{15pt minus 5pt}
\setlength{\dbltextfloatsep}{15pt minus 5pt}

\title{ResGraph: Cross-Model Graph-based Index Migration via Residual
Reachability in ANN Search}

\author{
  \IEEEauthorblockN{Baoyuan Gu\IEEEauthorrefmark{1},
    Xiaoyao Zhong\IEEEauthorrefmark{1},
    Jiabao Jin\IEEEauthorrefmark{1},
    Peng Cheng\IEEEauthorrefmark{1},
    Wangze Ni\IEEEauthorrefmark{2}\\
    Haoyang Li\IEEEauthorrefmark{4},
    Haotian Li\IEEEauthorrefmark{3},
    Jingkuan Song\IEEEauthorrefmark{1},
    Heng Tao Shen\IEEEauthorrefmark{1}}
  \IEEEauthorblockA{
    \IEEEauthorrefmark{1}Tongji University, Shanghai, China,
    \IEEEauthorrefmark{2}Zhejiang University, Hangzhou, China,\\
    \IEEEauthorrefmark{3}Ant Group, Shanghai, China,
    \IEEEauthorrefmark{4}Hong Kong Polytechnic University, Hong Kong SAR, China\\
    gu\_baoyuan@tongji.edu.cn; xiaoyao.zhong@hotmail.com;
    jiabaojin0723@gmail.com;\\
    cspcheng@tongji.edu.cn; niwangze@zju.edu.cn;
    haoyang-comp.li@polyu.edu.hk;\\ tianlan.lht@antgroup.com;
    jingkuan.song@gmail.com; shenhengtao@hotmail.com}
}

\maketitle

\begin{abstract}
Approximate nearest neighbor search (ANNS) underpins large-scale vector
retrieval in search, recommendation, and retrieval-augmented generation.
Graph-based indexes have demonstrated state-of-the-art search performance for
ANNS: they connect each corpus vector to a small set of nearby or
navigationally useful vertices and answer queries by traversing the resulting
graph.  Because these edges are selected using construction-time
distances, the graph index is tied to the embedding model; re-encoding a
corpus with a new model may change distances and neighborhoods of the
vectors, and reconstructing the graph for the new embedding vectors
incurs substantial construction cost and delays deployment.  When the embedding model changes, we observe a phenomenon
in the old graph index that we call \emph{residual reachability}: although
derived from different models, the vectors describe the same underlying
objects and often retain part of their similarity structure, which the old
graph reflects as connectivity, leaving many exact new-model neighbors
reachable within a few hops.  Motivated by this observation, we develop
ResGraph, an index-migration framework that exploits the residual reachability
of the old graph index to construct the new graph index faster.  ResGraph
provides two migration paths.
Res-Local performs parallel shallow expansion over the
inherited graph index and screens second-hop candidates with packed position sign
codes before exact evaluation.  Res-Search uses hop-bounded beam traversal to
explore beyond shallow expansion.
Across eight text and image migrations, ResGraph attains up to 17.43 times speedup on constructing the new graph index than the fastest degree-matched
reconstruction method while keeping competitive recalls.
\end{abstract}

\section{Introduction}\label{sec:intro}

Approximate nearest neighbor search (ANNS) supports dense retrieval,
recommendation, and retrieval-augmented generation~\cite{karpukhin2020dense,
	covington2016deep,lewis2020rag}.  Graph-based indexes have become a leading
paradigm for ANNS, offering a favorable trade-off between search accuracy and
efficiency~\cite{wang2021survey}.
Methods such as HNSW~\cite{malkov2020hnsw}, NSG~\cite{fu2019nsg}, and
Vamana~\cite{subramanya2019diskann} represent each vector as a vertex and
connect it to a small set of nearby or navigationally useful vertices.  Since
connecting edges are chosen based on distances between vectors, the
graph is specific to the embedding model producing the vectors.

\begin{figure}[t!]
		\centering
		\includegraphics[width=\columnwidth,trim=0 0 0 8.64pt,clip]{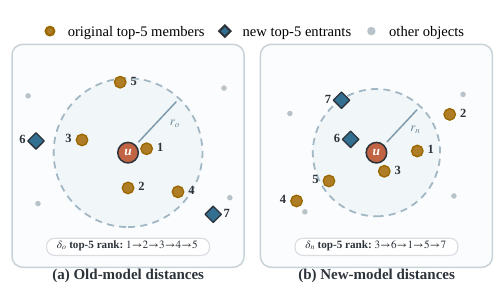}\vspace{-2ex}
		\caption{Neighborhood Drift after Embedding Model Replacement.}\vspace{-2ex}
		\label{fig:radius_ball}
\end{figure}

In deployed retrieval systems, embedding models evolve with advances in
model capabilities and changing application requirements; replacing an
encoder, often to improve retrieval quality or to serve a new domain,
re-encodes every corpus object (e.g., text or image).  Before a model is
deployed, several candidates are indexed over the same corpus and compared on
retrieval quality, after which the index of the selected model is built at
production scale.  To improve embedding quality and the final retrieval
quality, production systems adopt models better suited to the task, and each
adoption leads to substantial index reconstruction cost.

\noindent\textbf{Example 1 (Neighborhood drift after model replacement).}
Consider an object $u$ whose old-model top-5 neighbors, ordered by distance,
are $1 \rightarrow 2 \rightarrow 3 \rightarrow 4 \rightarrow 5$.  The
corresponding threshold radius is $r_\mathbf{o}=\delta_\mathbf{o}(u,5)$, where
$\delta_\mathbf{o}(u,5)$ denotes the distance between the embeddings of objects
with IDs $u$ and 5 in space $\mathbf{o}$.  After the new model re-encodes the
corpus into space $\mathbf{n}$, the top-5 order of $u$ becomes
$3 \rightarrow 6 \rightarrow 1 \rightarrow 5 \rightarrow 7$, with threshold
radius $r_\mathbf{n}=\delta_\mathbf{n}(u,7)$, as illustrated in \Cref{fig:radius_ball}.  Thus,
IDs 2 and 4 leave the top-5 neighborhood, IDs 6 and 7 enter it, and the shared
neighbors are reordered.

Re-encoding corpus objects updates the stored vectors but leaves the graph-index
edges unchanged.  Search therefore evaluates candidates using new-model
distances while traversing edges inherited from the old graph.  This
\textit{graph--space mismatch} can invalidate both local neighborhood relations
and the navigation cues used by graph search.

\begin{figure}[t!]
	\centering
	\includegraphics[width=\columnwidth]{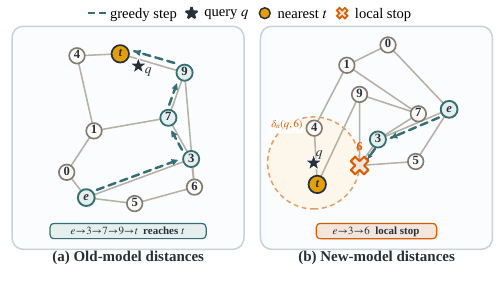}\vspace{-2ex}
	\caption{Graph Search Degradation under Graph--Space Mismatch.}\vspace{-2ex}
	\label{fig:overview}
\end{figure}

\noindent\textbf{Example 2 (A false local optimum under inherited edges).}
Consider the query $q$ and entry vertex $e$ in \Cref{fig:overview}.  Under
the old model, greedy search follows
$e \rightarrow 3 \rightarrow 7 \rightarrow 9 \rightarrow t$ and reaches the
true nearest neighbor $t$.  After re-encoding, evaluating the same stored
edges with new-model distances produces the route
$e \rightarrow 3 \rightarrow 6$.  At node 6, every stored neighbor lies
outside the query-centered ball of radius $\delta_\mathbf{n}(q,6)$, whereas $t$
lies inside it.  No stored edge therefore exposes a closer candidate, and the
search terminates at node 6 even though $t$ is substantially closer.

These examples expose two coupled challenges of embedding model replacement:
(1) \emph{graph--space mismatch}, where inherited edges can produce
incorrect local neighborhoods and premature search termination; and
(2) \emph{reconstruction overhead}, since the standard remedy must rediscover,
compare, and select candidate neighbors for every re-encoded vector,
substantially delaying the model-update process and motivating a natural
question: \textit{can the new-model index be constructed without incurring
the high cost of a full reconstruction?}

Existing update techniques make assumptions that do not hold in this setting.
Incremental maintenance keeps the embedding space fixed
~\cite{malkov2020hnsw,xu2023spfresh}, while compatibility methods constrain or
align the representations themselves
~\cite{shen2020bct,hu2022compatible,moschella2023relative,
	yang2025integrating,yoon2025converter}.  A model upgrade instead replaces
nearly every corpus vector and requires an index that serves the materialized
new embeddings directly.

To better understand how an embedding model change affects a graph-based ANNS
index, we distinguish between edge validity and candidate reachability.  When
evaluated using new-model distances, inherited edges may provide weaker
neighborhood relationships and navigation cues.  Nevertheless, the old and new
models encode the same ID-aligned objects and often preserve some of the
similarity relations among them, materialized in the old graph as both
direct adjacency and multi-hop connectivity.  We call the resulting empirical
phenomenon \emph{residual reachability}: an exact new-model neighbor may
disappear from a vertex's old adjacency list while remaining reachable
through a short old-graph path, leaving the old graph a rich pool of
candidates for new-model neighborhoods.
Across our eight migrations, exact new-model top-10
coverage rises from 9.7--35.3\% at one hop to 81.9--99.2\% within three hops.
At the same time, the degradation of inherited distance rankings
makes new-space distance evaluation essential.

These observations motivate our study of \emph{graph-index migration}:
transforming an existing graph index for a new embedding space by using residual
reachability in the old graph to accelerate construction of the new graph index.

\par\noindent\textit{Contributions.}\space
This paper makes three contributions:
\begin{itemize}[leftmargin=*,itemsep=1pt,topsep=2pt]
	\item We formalize graph-index migration, in which an old-space graph index is
	reused to accelerate construction of an index for the new space
	(\S\ref{sec:preliminaries}).
	\item We develop ResGraph for efficient graph-index
	migration.  Res-Local
	provides early availability through parallel
	shallow expansion, screening of second-hop candidates with packed sign codes, and exact
	new-space evaluation (\S\ref{sec:local_migration}).  Res-Search leverages the
	hop-distance distribution of new-model neighbors in the old graph and uses
	hop-bounded beam traversal to discover candidates dynamically.

	\item We evaluate ResGraph against four reconstruction baselines across eight
	migration scenarios and use screening, full-pool, and hop-ceiling ablations to
	characterize its construction efficiency and search performance
	(\S\ref{sec:experiments}).
\end{itemize}

\section{Problem Formulation}\label{sec:preliminaries}

\subsection{Basic Settings}\label{sec:problem}

Let $V=\{1,\ldots,n\}$ be the set of object IDs.  For an embedding space
$\mathbf{s}\in\{\mathbf{o},\mathbf{n}\}$, embedding model
$\mathbb M_\mathbf{s}$ maps each object $u\in V$ to a vector
$\mathbf x_u^\mathbf{s}\in\mathbb R^{d_\mathbf{s}}$.  We note that
$\mathbb M_\mathbf{s}$ defines a $d_\mathbf{s}$-dimensional space
$\mathbf{s}$.  The ID-paired vector
collection is $\mathcal X^\mathbf{s}=\{\mathbf x_u^\mathbf{s}\}_{u\in V}$.
For objects $u,v\in V$, let $\delta_\mathbf{s}(u,v)$ denote their distance in
space $\mathbf{s}$, and let $\mathcal N_k^\mathbf{s}(u)\subseteq
V\setminus\{u\}$ be the exact top-$k$ neighbor set of $u$.

\subsection{Graph-Based ANNS Indexes}\label{sec:graph_background}

A graph-based ANNS index organizes the vectors as a directed proximity graph
(PG)~\cite{malkov2020hnsw,fu2019nsg,subramanya2019diskann} to store the
similarity connections between the vectors.

\begin{definition}[Proximity Graph]
The PG of $\mathcal X^\mathbf{s}$ is
$G^\mathbf{s}=(\mathcal V^\mathbf{s},E^\mathbf{s})$, where
$\mathcal V^\mathbf{s}=\{v_u:u\in V\}$ and vertex $v_u$ stores
$\mathbf x_u^\mathbf{s}$.  A directed edge $(v_i,v_j)\in E^\mathbf{s}$
represents proximity under $\delta_\mathbf{s}$, evaluated by a specific
distance metric (e.g., cosine).  The adjacency lists of object $u$ is
$N_{G^\mathbf{s}}(u)=\{j\in V:(v_u,v_j)\in E^\mathbf{s}\}$,
with out-degree bounded as $|N_{G^\mathbf{s}}(u)|\le M$.
\end{definition}

At query time, graph search starts from one or more entry vertices and keeps a
bounded candidate pool.  It evaluates query-to-vertex distances and expands
adjacency lists.  The
parameter $\mathrm{ef}$ controls this pool and the accuracy--throughput
trade-off.  For a query $\mathbf q$ in space $\mathbf{s}$, let
$T_k(\mathbf q)$ and $\hat T_k(\mathbf q)$ denote the exact and returned
top-$k$ ID sets, respectively; the graph and $\mathrm{ef}$ setting are fixed
by context.  The search accuracy for $\mathbf q$ is
\begin{equation}
  \operatorname{Recall@}k(\mathbf q)
  =
  \frac{\left|T_k(\mathbf q)\cap\hat T_k(\mathbf q)\right|}{k}.
  \label{eq:recall}
\end{equation}
Reported recall is averaged over the test queries.

\begingroup
\setlength{\emergencystretch}{1em}
Most index constructions first acquire candidate neighbors through graph search,
neighbor joins, or another construction routine and then select a
degree-bounded subset~\cite{dong2011nndescent,wang2021survey}.   Candidate
availability  and  neighbor selection  jointly determine the final graph; ResGraph
reuses the former but recomputes the latter in the new space.\par
\endgroup

\subsection{Graph-Index Migration}\label{sec:migration_problem}

Let $G^\mathbf{o}=(\mathcal V^\mathbf{o},E^\mathbf{o})$ be the original
graph index built for $\mathcal X^\mathbf{o}$.  When a new embedding model
$\mathbb M_\mathbf{n}$ is deployed, we introduce a straightforward operation,
\emph{Res-Replace}, which replaces the stored vector of every object while
leaving the ID-paired edge relation unchanged.

\begin{definition}[Res-Replace]
\label{def:dgm_replace}
Given an original graph index $G^\mathbf{o}=(\mathcal{V}^\mathbf{o},E^\mathbf{o})$ and a new embedding model $\mathbb{M}_\mathbf{n}$, \emph{Res-Replace} generates a graph index $\tilde{G}^\mathbf{o}=(\mathcal{V}^\mathbf{n},E^\mathbf{o})$ through only replacing each object $u$'s original embedding $x_u^\mathbf{o}\in \mathcal{X}^\mathbf{o}$ with its new embedding $x_u^\mathbf{n}\in \mathcal{X}^\mathbf{n}$.
\end{definition}

Res-Replace thus exposes how inherited connectivity behaves under new-model
distances and provides the starting state for our residual-reachability
analysis; we use it as a diagnostic control that separates vector
replacement from graph adaptation.

\begin{definition}[Graph-index migration]
\label{def:graph_migration}
Given an original graph-based index $G^\mathbf{o}=(\mathcal{V}^\mathbf{o},E^\mathbf{o})$ and a new embedding model $\mathbb{M}_\mathbf{n}$,
graph-index migration constructs a graph
$\hat{G}^\mathbf{n}=(\mathcal{V}^\mathbf{n},\hat{E}^\mathbf{n})$ through reusing the information from $G^\mathbf{o}$.
\end{definition}

The goal of graph-index migration is to save the construction time needed to
adapt to the new embedding model after the new embedding vectors
$\mathcal X^\mathbf{n}$ have been materialized.  Full reconstruction builds
$G^\mathbf{n}=(\mathcal V^\mathbf{n},E^\mathbf{n})$ from
$\mathcal X^\mathbf{n}$ alone, whereas migration constructs
$\hat G^\mathbf{n}$ from $G^\mathbf{o}$ by exploiting its residual
reachability.

\Cref{tab:notation} summarizes the symbols used throughout the paper.

\begin{table}[t]
	\centering
	\caption{Frequently Used Symbols.}
	\label{tab:notation}
	\footnotesize
	\setlength{\tabcolsep}{3pt}
	\renewcommand{\arraystretch}{1.08}
	\begin{tabular}{@{}cl@{}}
		\toprule
		Symbol & Description \\
		\midrule
		$\mathbf{s}\in\{\mathbf{o},\mathbf{n}\}$ & Old or new embedding space. \\
		$\mathbb M_\mathbf{s},d_\mathbf{s}$ & Embedding model and dimension of space $\mathbf{s}$. \\
		$\mathbf x_u^\mathbf{s},\mathcal X^\mathbf{s}$ & Vector of object $u$ and the vector set of objects. \\
		$V,n,\mathcal V^\mathbf{s}$ & ID set, corpus size, and corresponding graph vertices. \\
		$w$ & Machine-word width used by packed sign codes. \\
		$k,\mathrm{ef}$ & Result size and query-time search parameter. \\
		$\delta_\mathbf{s}(u,v)$ & Distance between $u,v$ in space $\mathbf{s}$. \\
		$G^\mathbf{s}=(\mathcal V^\mathbf{s},E^\mathbf{s})$ & Proximity graph in space $\mathbf{s}$. \\
		$G^\mathbf{o},\tilde G^\mathbf{o}$ & Original and replace-only graph states. \\
		$G^\mathbf{n},\hat G^\mathbf{n}$ & Fully reconstructed and migrated new-model graphs. \\
		$\hat G^{(r)},R$ & Round-$r$ migrated graph and chain length. \\
		$N_G(u)$ & Out-neighbors of vertex $v_u$ in $G$, excluding $u$. \\
		$\mathcal N_k^\mathbf{s}(u)$ & Exact top-$k$ neighbors of $u$, excluding $u$. \\
		$T_k(\mathbf q),\hat T_k(\mathbf q)$ & Exact and returned top-$k$ sets of query $q$. \\
		$\mathcal P_{\alpha,M}$ & Diversity-aware neighbor-pruning rule. \\
		$\Gamma_h^\mathbf{o}(u)$ & Vertices outward-reachable within $h$ hops in $G^\mathbf{o}$. \\
		$\mathcal C_u,\mathcal Q_u$ & Second-hop candidate pool and MPS shortlist. \\
		$C$ & Maximum number of screened second-hop candidates. \\
		$b_u^{\mathrm{MPS}}$ & Mean-centered position sign code. \\
		$M,\alpha$ & Out-degree bound and pruning parameter. \\
		$L,H$ & Res-Search beam capacity and hop ceiling. \\
		$R_u,F_u$ & Search result pool and expansion frontier. \\
		\bottomrule
	\end{tabular}
\end{table}

\section{Residual Reachability: Empirical Motivation}
\label{sec:structural}

\begin{table}[!t]
  \caption{Structural Preservation across the Eight End-to-End Workloads.}
  \label{tab:structural_preservation}
  \centering
  \scriptsize
  \setlength{\tabcolsep}{3pt}
  \begin{tabular*}{\columnwidth}{@{\extracolsep{\fill}}lrrrrrr@{}}
    \toprule
    & \multicolumn{4}{c}{Exact top-10 coverage}
    & \multicolumn{2}{c}{Rank retention} \\
    \cmidrule(lr){2-5}\cmidrule(lr){6-7}
    Dataset & $\mathrm{cov}_1$ & $\mathrm{cov}_2$ & $\mathrm{cov}_3$ & $\mathrm{cov}_4$
    & $\rho$ & $\tau$ \\
    \midrule
    CIFAR-10    & .177 & .619 & .984 & 1.000 & .612 & .438 \\
    AG News     & .353 & .806 & .988 & 1.000 & .886 & .719 \\
    Yelp        & .141 & .419 & .819 & .980  & .587 & .420 \\
    Amazon-1536 & .334 & .736 & .984 & 1.000 & .793 & .606 \\
    FineWeb     & .205 & .601 & .949 & .998  & .702 & .517 \\
    Yahoo       & .327 & .755 & .992 & 1.000 & .748 & .558 \\
    QuickDraw   & .097 & .390 & .819 & .972  & .525 & .371 \\
    MS MARCO    & .323 & .763 & .980 & 1.000 & .841 & .662 \\
    \bottomrule
  \end{tabular*}
\end{table}

\begin{table}[!t]
  \caption{Scale-Matched Enrichment of Old-Graph Hop Sets.}
  \label{tab:hop_enrichment}
  \centering
  \scriptsize
  \setlength{\tabcolsep}{3pt}
  \begin{tabular*}{\columnwidth}{@{\extracolsep{\fill}}lrrrrrr@{}}
    \toprule
    & \multicolumn{2}{c}{$h=2$} & \multicolumn{2}{c}{$h=3$}
    & \multicolumn{2}{c}{$h=4$} \\
    \cmidrule(lr){2-3}\cmidrule(lr){4-5}\cmidrule(l){6-7}
    Dataset & $f_2$ & $E_2$ & $f_3$ & $E_3$ & $f_4$ & $E_4$ \\
    \midrule
    CIFAR-10    & 2.196\% & 28.17$\times$   & 28.99\% & 3.40$\times$   & 83.80\% & 1.19$\times$ \\
    AG News     & 0.660\% & 122.18$\times$  & 10.92\% & 9.05$\times$   & 59.64\% & 1.68$\times$ \\
    Yelp        & 0.151\% & 277.88$\times$  & 3.36\%  & 24.34$\times$  & 31.81\% & 3.08$\times$ \\
    Amazon-1536 & 0.098\% & 750.70$\times$  & 1.88\%  & 52.36$\times$  & 14.29\% & 7.00$\times$ \\
    FineWeb     & 0.181\% & 332.54$\times$  & 5.73\%  & 16.55$\times$  & 59.53\% & 1.68$\times$ \\
    Yahoo       & 0.199\% & 380.45$\times$  & 6.17\%  & 16.09$\times$  & 58.15\% & 1.72$\times$ \\
    QuickDraw   & 0.062\% & 625.95$\times$  & 1.16\%  & 70.86$\times$  & 10.76\% & 9.03$\times$ \\
    MS MARCO    & 0.018\% & 4150.12$\times$ & 0.60\%  & 164.39$\times$ & 10.30\% & 9.71$\times$ \\
    \bottomrule
  \end{tabular*}
\end{table}

Although embedding coordinates change across models, similarity relations
among the same objects may persist. Many exact new-model neighbors remain
reachable within a few hops in the old graph despite being absent from
its adjacency lists. We call this phenomenon \emph{residual reachability}.

\paragraph{Coverage of new-model neighbors.}
We measure residual reachability on the eight workloads in
Table~\ref{tab:datasets}, applying Res-Replace to $G^{\mathbf{o}}$ to
update its vectors while preserving its edges. For a vertex $u$,
let $\Gamma^{\mathbf{o}}_h(u)$ denote the vertices reachable from $u$
within $h$ hops along the outgoing edges of $G^{\mathbf{o}}$, excluding
$u$ itself. The fraction of exact new-model neighbors in this set is
\begin{equation}
  \operatorname{cov}_h(u)=
  \frac{\left|\mathcal{N}^{\mathbf{n}}_k(u)\cap
        \Gamma^{\mathbf{o}}_h(u)\right|}{k}.
  \label{eq:hop_coverage}
\end{equation}
We set $k=10$ and average the results over a set $\mathcal S$ of
128 sampled vertices, denoting the mean by
$\overline{\operatorname{cov}}_h$. Coverage is cumulative: the two-hop set includes direct and
second-hop neighbors.

Table~\ref{tab:structural_preservation} reports coverage and rank retention
across all eight workloads.

\paragraph{Size of the candidate sets.}
To distinguish coverage gains from candidate-set growth, we compare
each hop set with an equally sized uniform sample. We define
\begin{equation}
  f_h=\frac{1}{|\mathcal S|}\sum_{u\in\mathcal S}
      \frac{|\Gamma^{\mathbf{o}}_h(u)|}{n-1},
  \qquad
  E_h=\frac{\overline{\operatorname{cov}}_h}{f_h}.
  \label{eq:hop_enrichment}
\end{equation}
Here, $f_h$ is the mean relative size of the hop set. For each $u$, uniformly sampling
$|\Gamma^{\mathbf{o}}_h(u)|$ vertices from $V\setminus\{u\}$ gives an
expected top-10 coverage of $|\Gamma^{\mathbf{o}}_h(u)|/(n-1)$.
Averaging over $\mathcal S$ gives $f_h$. Thus, $E_h$ measures the
enrichment of the old-graph candidate sets relative to this uniform baseline.

Table~\ref{tab:hop_enrichment} reports candidate-set fractions of
0.02--2.20\%, 0.60--28.99\%, and 10.30--83.80\% at depths two,
three, and four, respectively, with corresponding enrichment of
28.17--4150.12$\times$, 3.40--164.39$\times$, and 1.19--9.71$\times$.
All reported two- to four-hop sets enrich exact new-model neighbors
over equally sized uniform samples.

\paragraph{Distance rankings.}
For 1,000 sources, the old- and new-distance rankings of
deduplicated second-hop candidates have mean Spearman's $\rho$ of
0.525--0.886 and Kendall's $\tau$ of 0.371--0.719
(Table~\ref{tab:structural_preservation}).

\begin{figure}[!t]
  \centering
  \includegraphics[width=\columnwidth]{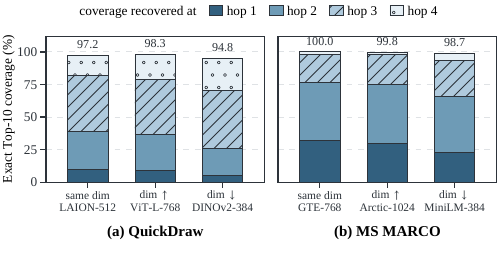}\vspace{-2ex}
  \caption{Residual Reachability under Dimensional Change.}\vspace{-2ex}
  \label{fig:dimension_changing_reachability}
\end{figure}

\paragraph{Changes in model architecture and dimension.}
On QuickDraw and MS MARCO, whose old graphs are built with MetaCLIP-B/32-512
and e5-base-768 respectively, we compare architecture- and dimension-changing
replacements with same-dimensional controls, fixing the old graph and 128
anchor vertices across targets within each dataset.
Figure~\ref{fig:dimension_changing_reachability} stacks, for each target, the
exact new-model top-10 coverage first recovered at old-graph hop $h$; the bar
height is therefore the coverage recovered within four hops.
\emph{Residual reachability persists under changes in both model
architecture and embedding dimension}: short old-graph paths recover
nearly all exact new-model neighbors by four hops.

These observations motivate screened two-hop expansion in Res-Local
(\S\ref{sec:local_migration}) and hop-bounded beam search in Res-Search
(\S\ref{sec:pipeline}).

\section{Res-Local: Migration via Screened Local Expansion}
\label{sec:local_migration}

Residual reachability persists across datasets and model replacements.
Res-Local therefore retrieves many new-model neighbors from the old graph
through short $h$-hop expansion, accelerates candidate screening with a
packed sign code, and re-checks the shortlist with exact distances.

\subsection{Packed Sign Screening}
\label{sec:candidate_evidence}
\label{sec:dual_sign_repair}

Res-Local turns residual reachability into a bounded pool for exact evaluation
in two stages.  It first expands the old graph to expose promising candidates
and then applies a lightweight packed sign screen to retain a bounded
shortlist.  Specifically, the second hop exposes many new-model neighbors
without requiring a deeper graph traversal.  Res-Local therefore collects, for
every vertex $u$, the deduplicated second-hop candidate set
\begin{equation}
 \mathcal C_u=\Gamma_2^\mathbf{o}(u)\setminus
   \bigl(N_{G^\mathbf{o}}(u)\cup\{u\}\bigr).
 \label{eq:local_pool}
\end{equation}
Extending this static expansion further would perform a deeper traversal and
produce a pool that grows quickly with graph degree.

Res-Local deduplicates the pool before scoring, ensuring that each candidate is
considered once.
\Cref{eq:local_pool} excludes inherited 1-hop neighbors because they bypass
screening and always receive an exact new-model distance evaluation.  The
screen therefore controls only candidates introduced by the second hop.  Even
after deduplication, however, this pool may be much larger than the final
adjacency list.  Evaluating it directly would incur a high-dimensional distance
for every candidate and increase the work of diversity pruning.  We then propose packed sign screening to reduce $\mathcal C_u$ to a coarse shortlist before
any exact evaluation of its members.

Compact quantization codes are commonly used to screen candidates before exact
distance evaluation.  RaBitQ provides an unbiased distance estimator with a
sharp, asymptotically optimal error bound for one-bit-per-dimension
quantization~\cite{gao2024rabitq}, but its normalization, rotation, encoding,
and metadata are amortized across repeated queries.  Res-Local screens each
already
formed 2-hop pool only once and reranks a generous shortlist by exact
distances.  We therefore adopt a simpler temporary sign encoding scored by
Hamming agreement.  It uses a uniform calibration sample: given budget
$m_{\mathrm{cal}}$, let $S\subseteq V$ with
$|S|=\min\{m_{\mathrm{cal}},n\}$.  Vector inequalities and indicator
functions below are evaluated coordinatewise.

\paragraph{Mean-centered position sign code (MPS)}
It only describes the new space $\mathbf{n}$.  Let
$\bar{\mathbf x}^\mathbf{n}=\frac{\sum_{j\in S}\mathbf x_j^\mathbf{n}}{|S|}$; then
\begin{equation}
 b_u^{\mathrm{MPS}}=\mathbf{1}[\mathbf x_u^\mathbf{n}\ge\bar{\mathbf x}^\mathbf{n}].
 \label{eq:mps}
\end{equation}
It records the orthant of $u$ relative to a sampled center in the new
embedding space $\mathbf{n}$.  For two objects $u$ and $v$, the Hamming agreement is
\begin{equation}
 \sigma_{\mathrm{MPS}}(u,v)
 =\sum_{i=1}^{d_\mathbf{n}}\mathbf{1}[b_{u,i}^{\mathrm{MPS}}=b_{v,i}^{\mathrm{MPS}}]
 =d_\mathbf{n}-d_H(b_u^{\mathrm{MPS}},b_v^{\mathrm{MPS}}),
 \label{eq:sign_score}
\end{equation}
where $d_H$ denotes Hamming distance.
The construction extends
coordinatewise and is compared through \Cref{eq:sign_score}.
MPS induces a shortlist $\operatorname{Top}_C^{\mathrm{MPS}}(u)$ containing the
$\min(C,|\mathcal C_u|)$ objects with largest $\sigma_{\mathrm{MPS}}(u,v)$, using object ID as the
secondary order.  A partial selection avoids sorting the full pool.

The screening ablation in \S\ref{sec:exp:screening_ablation} shows that MPS
combines a simple construction with strong screening accuracy; full RaBitQ
improves coverage but incurs higher preparation, storage, and scoring
costs.  MPS is deliberately a lightweight heuristic rather than a distance
estimator: it controls only candidate admission, and the
admitted candidates are reranked by exact new-model distance.  Res-Local
therefore adopts MPS because it requires only the new space and better
matches a temporary, one-use coarse screen.

Res-Local therefore sets $\mathcal Q_u=\operatorname{Top}_C^{\mathrm{MPS}}(u)$.
Screening uses only packed integer data: the MPS array occupies
$\lceil d_\mathbf{n}/w\rceil$ machine words per object, and each comparison
evaluates the Hamming distance word by word.  Res-Local stores it by the IDs of objects
during migration and discards it afterward.  At $d_\mathbf{n}=768$ and $w=64$,
the code uses 96
bytes per object, compared with 3072 bytes for one 32-bit floating-point
(FP32) vector.

\subsection{Exact Distance Evaluation and Neighbor Pruning}

Res-Local unions the shortlist with all
inherited 1-hop neighbors.  It then computes the exact new model distance
from $u$ to every candidate in this union:
\begin{equation}
 \mathcal E_u=\{(\delta_\mathbf{n}(u,v),v):
 v\in\mathcal Q_u\cup N_{G^\mathbf{o}}(u)\}.
 \label{eq:exact_candidates}
\end{equation}
We write $\mathcal P_{\alpha,M}$ for the diversity-aware neighbor-pruning
rule used by the
graph implementation.  Applying it to $\mathcal E_u$
selects at most $M$ outgoing neighbors:
\begin{equation}
 N_{\hat G^\mathbf{n}}(u)=\mathcal P_{\alpha,M}(u,\mathcal E_u).
 \label{eq:local_pruned_neighbors}
\end{equation}
The screening code therefore governs admission to $\mathcal E_u$, not final edge
selection.  Exact new model distances order the evaluated candidates, and
diversity pruning determines the stored list.

\begingroup
\SetAlgoSkip{}
\begin{algorithm}[!t]
\caption{Res-Local for Vertex $u$}
\label{alg:dual_sign_refine}
\KwIn{source $u$; old graph $G^\mathbf{o}$; new embeddings $\mathcal X^\mathbf{n}$;
  packed codes $\{b_v^{\mathrm{MPS}}\}_{v\in V}$; shortlist cap $C$; out-degree bound
  $M$; pruning factor $\alpha$}
\KwOut{migrated neighbor list $N_{\hat G^\mathbf{n}}(u)$}
$\mathcal O_u\leftarrow N_{G^\mathbf{o}}(u)$\;
$\mathcal C_u\leftarrow\emptyset$\;
\ForEach{$z\in\mathcal O_u$}{
  insert every ID in $N_{G^\mathbf{o}}(z)$ into $\mathcal C_u$\;
}
remove $u$ and every ID in $\mathcal O_u$ from $\mathcal C_u$\;
deduplicate $\mathcal C_u$ by ID\;
$\mathcal S_u\leftarrow\emptyset$\;
\ForEach{$v\in\mathcal C_u$}{
  $s_v\leftarrow\sigma_{\mathrm{MPS}}(u,v)$\;
  append $(s_v,v)$ to $\mathcal S_u$\;
}
$k_u\leftarrow\min(C,|\mathcal C_u|)$\;
partially select the $k_u$ entries of $\mathcal S_u$ with largest scores\;
break equal-score ties by ID\;
$\mathcal Q_u\leftarrow$ the IDs in the selected entries\;
$\mathcal R_u\leftarrow\mathcal Q_u\cup\mathcal O_u$\;
$\mathcal E_u\leftarrow\emptyset$\;
\ForEach{$v\in\mathcal R_u$}{
  $d_v\leftarrow\delta_\mathbf{n}(u,v)$\;
  append $(d_v,v)$ to $\mathcal E_u$\;
}
$N_{\hat G^\mathbf{n}}(u)\leftarrow\mathcal P_{\alpha,M}(u,\mathcal E_u)$\;
\KwRet $N_{\hat G^\mathbf{n}}(u)$\;
\end{algorithm}
\endgroup

\par\noindent\textit{Res-Local algorithm.}\space
\Cref{alg:dual_sign_refine} records the inherited 1-hop neighbors, expands
their adjacency lists into a deduplicated 2-hop candidate pool, and removes
$u$ and the 1-hop neighbors before screening.  It scores each remaining
candidate by MPS Hamming agreement, retains the best
$\min(C,|\mathcal C_u|)$ by partial selection with ID tie-breaking, and
merges this shortlist with the inherited 1-hop neighbors into an
exact-evaluation pool, every member of which receives its exact
new-model distance: old 1-hop neighbors thus bypass the binary gate but
not exact evaluation.  Finally, diversity pruning selects at most $M$
outgoing neighbors from the evaluated candidates.

For a hierarchical input index, Res-Local preserves each object's level and the inherited
entry point.  It repeats candidate generation, new model distance evaluation,
and pruning in every stored upper layer.  

\par\noindent\textit{Parallel execution and cost.}\space
Every thread reads the same old graph and writes a different output list;
consequently, vertex order does not affect Res-Local.  Each thread reuses its
candidate, score, and distance buffers.  Let $b_\mathbf{n}=\lceil d_\mathbf{n}/w\rceil$ be the
number of machine words per code.  Binary scoring at vertex $u$ then costs
$O(|\mathcal C_u|b_\mathbf{n})$ integer operations.  If
$m_u=|\mathcal Q_u\cup N_{G^\mathbf{o}}(u)|$, the method computes $m_u$ exact
new model distances from $u$ before pruning.  Temporary storage consists of
one corpus-level MPS array plus thread-local buffers for candidate
scoring and partial selection, so 2-hop pools are not materialized for
all vertices at once, no candidate pool is written to disk, and no exact
distance is repeated within a deduplicated pool.

\section{Res-Search: Migration Beyond Shallow Expansion}
\label{sec:pipeline}

Res-Local trades candidate scope for early availability: its inherited 2-hop
view supports an inexpensive parallel pass but cannot recover a useful
new model neighbor outside that range.  Res-Search, a separate
quality-oriented migration method, instead traverses the current graph
to discover such candidates dynamically, starting from the Res-Replace
or Res-Local graph.  For every object $u$, it
evaluates discovered candidates using exact new model distances and applies
diversity pruning to replace its adjacency list; the central challenge is
controlling
candidate-discovery computation when this traversal is repeated for all
$n$ corpus objects.

\subsection{Beam Search under Graph-Index Migration}
\label{sec:pipe:beam_limit}

Standard beam search maintains a bounded result pool $R$, a frontier
$F\subseteq R$ of retained but unexpanded candidates, and a visited set that
prevents repeated distance evaluation~\cite{malkov2020hnsw,wang2021survey}.
For expansion point $u$, new model distance $\delta_\mathbf{n}(u,v)$ orders the
candidates.  The closest frontier candidate is expanded, and its unseen
neighbors compete for the best $L$ positions in $R$; the surviving unexpanded
candidates remain in $F$.

In graph-index migration, this search runs once for every corpus object.  Beam
capacity $L$ bounds the number of retained candidates, but not the number of
hops traversed: while newly discovered vertices remain among the best $L$, the
frontier can continue along a multi-hop route.  Thus, even a narrow beam may
read distant adjacency lists and perform exact new-model distance evaluations
for each of the $n$ traversals.

Per-vertex beam search underpins the refinement pass in
Vamana- and NSG-style index
construction~\cite{subramanya2019diskann,fu2019nsg}, using the
target distance metric to guide traversal until the candidate
frontier empties.  While the inherited graph preserves residual
reachability after embedding replacement, standard greedy search
remains agnostic to this locality prior---traversing well beyond the
hop radius where the exact new-model neighbors concentrate.
Res-Search tailors the refinement pass to leverage this property:
(1) each beam is seeded directly from the vertex's inherited
adjacency list; (2) a single pass suffices due to the inherited
graph's high hop coverage (as analyzed in
\S\ref{sec:structural}); and (3) the hop ceiling $H$ converts this
coverage property into an explicit per-vertex work bound.  Removing
this ceiling recovers unbounded single-pass refinement; we quantify
the trade-offs of both variants in \S\ref{sec:exp:hop_ablation}.

\begin{figure}[!t]
  \centering
  \includegraphics[width=\columnwidth]{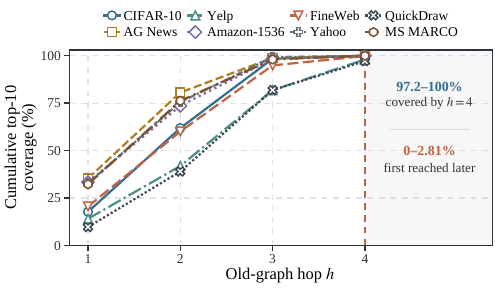}\vspace{-1ex}
  \caption{Exact Neighbor Concentration for the New Model.}\vspace{-1ex}
  \label{fig:hop_concentration}
\end{figure}

The structural measurements in \S\ref{sec:structural} show that most exact
new-model neighbors remain within a few old-graph hops.
Figure~\ref{fig:hop_concentration} plots cumulative exact new model top-10 coverage
against old-graph hop, averaged over 128 sampled vertices per workload.  Across
the eight workloads, three hops cover 81.9\%--99.2\% of these neighbors and
four cover 97.19\%--100\%, yet standard beam search does not use these
small hops as a stopping opportunity, so Res-Search adds a
hop ceiling $H$: a candidate discovered at depth $H$ remains eligible for the
result pool but is not expanded.  The two constraints are complementary:
$L$ bounds the retained search width, whereas $H$ bounds propagation depth.

\subsection{Hop-Ceiling Selection and Search Behavior}\label{sec:pipe:search}

Figure~\ref{fig:hop_concentration} shows coverage saturating within three
to four hops for most model pairs.  We compare these workload-specific
bounds with unbounded references in \S\ref{sec:exp:hop_ablation}.

To expose sensitivity to the hop ceiling rather than only the selected
operating points, we test $H\in\{3,4,5\}$ on Amazon-1536, QuickDraw, and
MS~MARCO with all other refinement parameters fixed: tighter ceilings
save more core refinement time but can discard useful propagation
routes, and quality approaches the unbounded result as $H$ grows.  We
therefore adopt $H=4$ ($H=3$ on CIFAR-10 and AG News), which cuts core
refinement time by 17.9--43.7\% across the eight workloads while
changing mean Recall@10 over the shared $\mathrm{ef}$ values by at
most $0.99$\,pp relative to unbounded expansion
(\S\ref{sec:exp:hop_ablation}).

\paragraph{Starting graph.}
Res-Local is not a prerequisite for Res-Search: starting from the current
neighbors, Res-Search revisits the shallow region covered by the local
expansion and then propagates through its retained frontier to more distant
vertices, so using the Res-Local graph mainly changes the initial shallow
connections rather than the region ultimately explored.  A controlled
starting-graph ablation illustrates this overlap: at query $\mathrm{ef}=50$,
starting from Res-Replace and Res-Local yields Recall@10 of 0.9854 and 0.9866
on Amazon-1536, and 0.9882 and 0.9896 on QuickDraw---gains of
only 0.12 and 0.14 percentage points.
Optional Res-Local contributes 11.7\% and 23.7\%
of the respective combined Res-Local and Res-Search time.  Thus,
Res-Search alone may start from Res-Replace; Res-Local is useful when an
inexpensive intermediate index is needed.

\setcounter{topnumber}{1}
\begin{figure}[!t]
	\centering
	\makebox[\columnwidth][c]{%
		\includegraphics[width=\columnwidth]{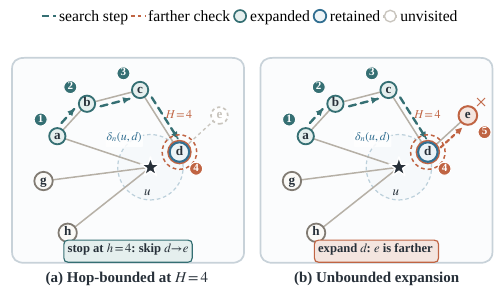}%
	}\vspace{-1ex}
	\caption{Effect of a Four-Hop Ceiling on an $L=1$ Search.}\vspace{-1ex}
	\label{fig:hop_boundary}
\end{figure}

For each expansion point $u$, Res-Search seeds the beam with $u$'s current
neighbors at depth one, or the global entry at depth zero when the list is
empty; \S\ref{sec:exp:hop_ablation} isolates how much this seeding choice
contributes.  The result pool $R_u$ and frontier $F_u$ retain their standard
beam-search roles, with $F_u\subseteq R_u$.  Candidate keys
$(\delta_\mathbf{n}(u,v),v)$ are ordered lexicographically by new-model
distance and then object ID.  Each candidate is evaluated only on first
discovery and assigned that discovery depth $h_u(v)$.

The hop constraint changes frontier admission, not result admission.  A
candidate with $h_u(v)\le H$ may enter $R_u$ and ultimately become a stored
neighbor, but it enters $F_u$ only when $h_u(v)<H$.  Hence a depth-$H$
candidate is evaluated and remains eligible for neighbor selection, while its
adjacency list is not read.  \Cref{fig:hop_boundary} uses the same synthetic
coordinates for source $u$ and its one-hop candidates $a,g,h$ in both
panels.  With $L=1$, both searches follow the same distance-ordered route to
hop four and retain $d$.  Only the unbounded search then reads
$N_G(d)$, evaluates the farther hop-five candidate $e$, and rejects it.
The bounded search avoids that adjacency-list read and exact distance
evaluation without changing $R_u$.
The full comparison with unbounded refinement, including the avoided work and
its recall effects, is reported in \S\ref{sec:exp:hop_ablation}.

\begin{figure*}[!t]
	\centering
	\includegraphics[width=\textwidth]{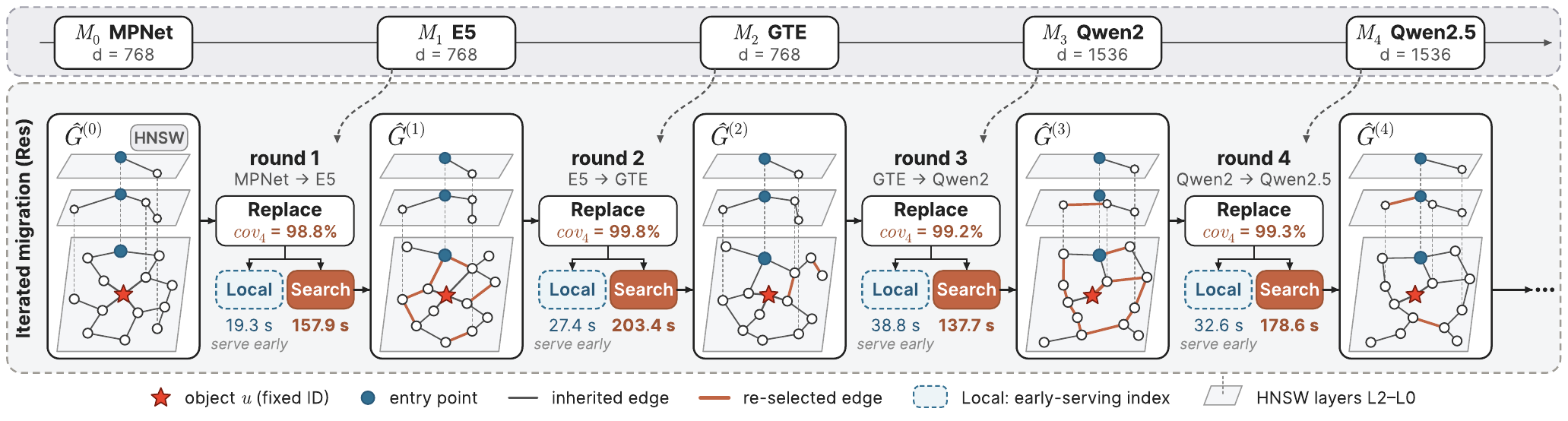}\vspace{-1ex}
	\caption{Iterated Migration across Model Releases.}\vspace{-1ex}
	\label{fig:iteration_scenario}
\end{figure*}

\begin{algorithm}[!t]
\caption{Hop-Bounded Migration Search for Expansion Point $u$}
\label{alg:hop_beam}
\KwIn{expansion point $u$; graph $G$; new-model vectors $\mathcal X^\mathbf{n}$;
  entry $e$; bounds $L,H$; pruning parameters $M,\alpha$}
\KwOut{updated new-model neighbor list $N_G(u)$}
initialize priority sets $R_u,F_u$ and
  $\mathrm{Seen}_u\leftarrow\{u\}$\;
$\mathcal I_u\leftarrow N_G(u)\setminus\{u\}$; set $h_u(z)\leftarrow1$
  for $z\in\mathcal I_u$\;
\If{$\mathcal I_u=\emptyset$ and $e\neq u$}{
  $\mathcal I_u\leftarrow\{e\}$; $h_u(e)\leftarrow0$\;
}
\ForEach{$z\in\mathcal I_u$}{
  add $z$ to $\mathrm{Seen}_u$; evaluate
    $d_z\leftarrow\delta_\mathbf{n}(u,z)$\;
  insert $(d_z,z)$ into $R_u$\;
  \If{$h_u(z)<H$}{
    insert $(d_z,z)$ into $F_u$\;
  }
  retain the best $L$ keys in $R_u$; set $F_u\leftarrow F_u\cap R_u$\;
}
\While{$F_u\neq\emptyset$}{
  pop the closest $v$ from $F_u$\;
  \ForEach{$z\in N_G(v)\setminus\mathrm{Seen}_u$}{
    add $z$ to $\mathrm{Seen}_u$; set
      $h_u(z)\leftarrow h_u(v)+1$\;
    evaluate $d_z\leftarrow\delta_\mathbf{n}(u,z)$\;
    \If{$|R_u|<L$ or $(d_z,z)$ precedes the worst key in $R_u$}{
      insert $(d_z,z)$ into $R_u$\;
      \If{$h_u(z)<H$}{
        insert $(d_z,z)$ into $F_u$\;
      }
      retain the best $L$ keys in $R_u$;
        set $F_u\leftarrow F_u\cap R_u$\;
    }
  }
}
$N_G(u)\leftarrow\mathcal P_{\alpha,M}(u,R_u)$\;
\KwRet $N_G(u)$\;
\end{algorithm}

\Cref{alg:hop_beam} initializes $R_u$, $F_u$, and $\mathrm{Seen}_u$
(Line~1) and seeds the traversal from $u$'s neighbors at depth one, or the
global entry at depth zero if the list is empty (Lines~2--4).  Each seed
is evaluated and inserted into $R_u$, only seeds below the ceiling enter
$F_u$, and beam truncation maintains $F_u\subseteq R_u$ (Lines~5--10).
Lines~11--12 expand the closest frontier vertex; on first discovery,
Res-Search records the depth and evaluates the new-model distance exactly once
(Lines~13--15).  The beam test controls admission to $R_u$, the depth
test admission to $F_u$ (Lines~16--20), so a vertex first reached at
depth $H$ may still become a stored neighbor although its edges are not
read.  Diversity pruning finally converts the retained pool into at most
$M$ outgoing edges (Line~21).

Each expansion point is processed once; the selected list replaces
$N_G(u)$, so later traversals already operate on the migrated
graph.  If the traversal for $u$ expands $a_u$ adjacency lists, it evaluates
at most $|N_G(u)|+Ma_u+1$ exact new-model source-to-candidate distances,
where the final term covers the optional global entry.  The hop ceiling
bounds traversal depth, beam pruning reduces the number of expanded lists,
and deduplication makes each candidate evaluated at most once, so residual
reachability is converted into per-vertex bounded work.
\setcounter{topnumber}{2}

\subsection{Composability of Iterated Migration}
\label{sec:iterated_migration}

A single migration prices one upgrade.  Deployment,
however, composes such upgrades: new releases of the embedding model keep
arriving, and once the migration to $\mathbb M_r$ finishes, release
$\mathbb M_{r+1}$ finds only the migrated graph as a legacy input.
Writing $\hat G^{(0)}=G^{\mathbf{o}}$, round $r$ transforms the serving
pair $(\hat G^{(r-1)},\mathbb M_r)$ into $\hat G^{(r)}$, composing $R$
successive migrations,
\begin{equation}
 \hat G^{(r)}=\operatorname{migrate}\bigl(\hat G^{(r-1)},\mathbb M_r\bigr),
 \qquad r=1,\ldots,R,
 \label{eq:iterated_migration}
\end{equation}
rather than a one-shot replacement.  \Cref{fig:iteration_scenario}
instantiates this composition on the four-round Amazon chain of
\S\ref{sec:exp:iteration}: each migrated graph supplies the next round's
legacy input.
Iteration adds two concerns that a single migration cannot expose.
First, residual reachability must be re-established at every round by a
graph whose edges were already re-selected for the previous model; were
the phenomenon a transient artifact of the original construction, it
would decay as migrations accumulate.  Second, per-round quality could
compound: any migration deficit relative to a fresh rebuild, however
small, may accumulate over $R$ rounds.  Both concerns are empirical;
\S\ref{sec:exp:iteration} evaluates two ID-aligned chains of released
models and finds that neither degradation occurs.

\section{Experimental Evaluation}\label{sec:experiments}

We evaluate ResGraph through four research questions (RQs).  \textbf{RQ1:} How quickly does
Res-Local provide a high-quality new-model index relative to complete
reconstruction?  \textbf{RQ2:} How effectively does Res-Search approach
reconstruction quality while preserving the construction advantage?
\textbf{RQ3:} Are these construction gains consistent across migrations with
different corpus sizes, dimensions, and model pairs, including target models
that change the embedding dimension?  \textbf{RQ4:} Which
design choices produce these gains: migrating adjacency beyond
replacement-only serving, MPS candidate screening, and hop-bounded
refinement?  After a common protocol, we report the overall performance of
Res-Local and Res-Search, including cross-migration evidence and
multi-round model iterations, and then isolate each mechanism through
ablations.

\subsection{Experimental Methodology}\label{sec:exp:setup}

\paragraph{Workloads.}
Each workload pairs old and new embedding models over the same object
set.  We use eight image and text migrations from 50K to 8.84M base
vectors and 384 to 4096 dimensions (\Cref{tab:datasets}), with 1,000 held-out
objects encoded by the new model as queries and exact offline top-10
neighbors to measure recall.

CIFAR-10 is the standard image corpus~\cite{krizhevsky2009cifar}, and
QuickDraw supplies the large sketch-image migration~\cite{ha2018neural}.
AG News, Yelp, Yahoo, and Amazon come from the text-classification
collections of Zhang et al.~\cite{zhang2015charcnn}; FineWeb uses the
curated web collection~\cite{penedo2024fineweb}; and MS\,MARCO is the
passage-retrieval benchmark~\cite{nguyen2016msmarco}.  The encoders cover
ResNet~\cite{he2016resnet}, CLIP-style vision encoders~\cite{radford2021clip},
MiniLM and
MPNet~\cite{wang2020minilm,song2020mpnet}, BGE~\cite{xiao2024cpack},
NV-Embed~\cite{lee2025nvembed}, and the Qwen2 and Qwen2.5
families~\cite{yang2024qwen2,qwen2024qwen25}, all as released checkpoints
without additional training.

\setcounter{dbltopnumber}{1}
\begin{table}[!t]
	\centering
	\caption{Main End-to-End Migration Workloads.}
	\label{tab:datasets}
	\ExpTableFont
	\setlength{\tabcolsep}{4.0pt}
	\begin{tabular}{@{}ccccc@{}}
		\toprule
		Dataset & $n$ & $d_\mathbf{o}=d_\mathbf{n}$ & Old model & New model \\
		\midrule
		CIFAR-10 & 50K & 512 & ResNet-18 & ResNet-34 \\
		AG News & 99K & 4096 & NV-Embed-v2 & SFR-Embedding-Mistral \\
		Yelp & 649K & 768 & DistilRoBERTa & BGE-base \\
		Amazon-1536 & 999K & 1536 & Qwen2-1.5B & Qwen2.5-1.5B \\
		FineWeb & 999K & 384 & Multi-QA-MiniLM & Paraphrase-MiniLM \\
		Yahoo & 999K & 768 & e5-base & gte-base \\
		QuickDraw & 1.28M & 512 & MetaCLIP-B/32 & LAION2B-CLIP-B/32 \\
		MS\,MARCO & 8.84M & 768 & e5-base & gte-base \\
		\bottomrule
	\end{tabular}
\end{table}

We additionally report a separate C4 English stress
test~\cite{raffel2020t5}: 99,999,000 base vectors and 1,000 held-out
queries for an e5-base-v2-to-gte-base migration at 768 dimensions,
analyzed separately in \S\ref{sec:exp:c4_100m} because it uses a narrower
baseline set.

Four additional dimension-changing workloads reuse the QuickDraw and
MS~MARCO original graphs: QuickDraw replaces MetaCLIP-B/32 (512) with
CLIP-ViT-L/14 (768) or DINOv2 (384); MS~MARCO replaces e5-base (768)
with Arctic (1,024) or MARCO-MiniLM-L12 (384).  All four use the same
1,000-query ground-truth/$\mathrm{ef}$ protocol, $M=64$, and 32 threads.

\paragraph{Compared methods and configurations.}
HNSW, NN-Descent, and the other graph baselines are implemented on top of the
open-source VSAG library~\cite{zhong2025vsag}.
The original graph has bottom-layer degree $M=64$ and is built with
construction pool $\mathrm{ef}_c=400$.  For the replacement and availability
protocols, Res-Local expands 2-hop on old-graph  and uses
$C=256$ below 500K vectors and $C=400$ otherwise.  The reported Res-Local
time includes migration of the bottom and upper layers.  For the refinement
comparison,
Res-Search starts directly from Res-Replace and processes every vertex once
with hop ceiling $H=3$ for CIFAR-10 and AG News
and $H=4$ otherwise, with beam capacity
$L=128$ for CIFAR-10 and AG News, $L=200$ for Yelp, Amazon-1536,
FineWeb, Yahoo, and QuickDraw, and $L=256$ for MS~MARCO.

The C4-100M stress test retains $M=64$, 32 threads, HNSW
$\mathrm{ef}_c=400$, and the main large-workload ResGraph settings (MPS with
$C=400$; $H=4$, $L=256$).

We compare against four degree-matched reconstruction baselines with their
distinct construction rules: HNSW (construction width 400, MRNG-style
relative-neighborhood selector), Vamana (width-400 pool, standard
RobustPrune), NSG (30-iteration NN-Descent graph, one width-400
search--collect pass, MRNG selection), and standalone NN-Descent (30
iterations).  A graph-reuse baseline seeds NN-Descent with the unmodified
old-graph neighbors, evaluates candidates in the new space, and runs ten
rounds at sample rate 0.2; ResGraph Replace (\Cref{def:dgm_replace}) is an
internal design control, not an existing baseline.  The
\emph{availability protocol} compares Res-Local with full
reconstructions, reporting time to a searchable graph and endpoint recall
relative to the fastest reconstruction baseline.  The \emph{quality
protocol} compares Res-Search with the same reconstruction baselines.

\paragraph{Metrics and measurement.}
All similarity computations use cosine distance over FP32 vectors.
Graph-processing time measures the interval until the new-model graph is ready for
queries.  Common embedding encoding and dataset loading time are excluded.
Search quality is Recall@10, averaged over the 1,000 held-out queries, and
throughput is single-threaded QPS.  We sweep query
$\mathrm{ef}\in\{10,20,30,40,50,100,150,200,300,400\}$ and connect the
measured nondominated points to obtain each Recall--QPS frontier.  The
mechanism studies additionally count exact new-model distance evaluations,
adjacency expansions, and traversal-distance evaluations.

Experiments run on a two-socket AMD EPYC 9334 server with 64 physical cores,
440\,GiB of memory, and local NVMe storage.  Every method uses 32 graph
threads, and the runtime dispatcher selects AVX-512 throughout.

\begin{table}[!t]
  \centering
  \caption{Time Advantage and Quality Cost of Res-Local.}
  \label{tab:local_tradeoff}
  \ExpTableFont
  \begin{tabular}{@{}cccc@{}}
    \toprule
    & \multicolumn{2}{c}{Time advantage ($\times$)} & Recall@10 loss \\
    \cmidrule(lr){2-3}
    Dataset & Fastest & Slowest & (pp) \\
    \midrule
    CIFAR-10    & 10.81 & 41.82 & 0.03 \\
    AG News     & 11.84 & 20.96 & 0.00 \\
    Yelp        & 17.10 & 42.93 & 1.49 \\
    Amazon-1536 & 13.37 & 39.24 & 0.06 \\
    FineWeb     & 11.76 & 39.38 & 0.43 \\
    Yahoo       & 14.20 & 26.16 & 0.00 \\
    QuickDraw   & 10.70 & 27.13 & 0.93 \\
    MS\,MARCO   & 17.43 & 40.27 & 0.06 \\
    \bottomrule
  \end{tabular}
\end{table}

\subsection{Overall Performance of Res-Local}\label{sec:exp:speed}

\subsubsection{Time to a Searchable New-Model Graph}
\label{sec:exp:local_quality}

\Cref{tab:local_tradeoff} answers RQ1 under a degree- and thread-matched
protocol.  The table reports full-reconstruction time relative to Res-Local
for the fastest and slowest baseline on each workload and the absolute
Recall@10 loss at the highest-recall measured point relative to the fastest
reconstruction.  Res-Local becomes searchable
$10.70\times$--$17.43\times$ sooner than the fastest full reconstruction baseline and
$20.96\times$--$42.93\times$ sooner than the slowest.  At the highest-recall
measured point, its Recall@10 loss relative to the same fastest reconstruction baseline is
0.00--1.49\,pp.  Six workloads lose at most 0.43\,pp.

\subsubsection{Res-Local versus Res-Replace}
\label{sec:exp:replace_control}

Res-Replace (\Cref{def:dgm_replace}) changes the stored vectors while retaining
the original edge relation.  Comparing it with Res-Local isolates the value of
migrating those lists.

\begin{figure}[!t]
  \centering
  \includegraphics[width=\columnwidth]{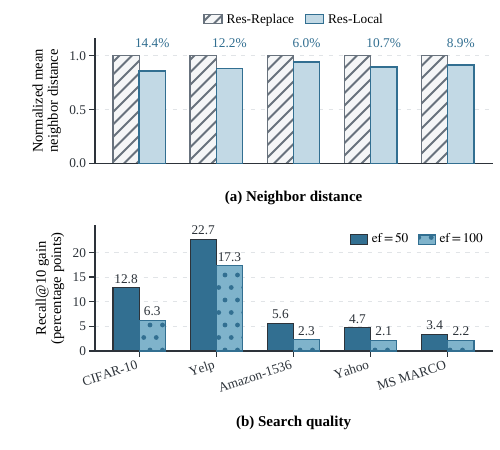}\vspace{-1ex}
  \caption{Effect of Adjacency-List Migration beyond Res-Replace.}\vspace{-1ex}
  \label{fig:local_vs_replace}
\end{figure}

We compare the designs on five workloads sharing the same old
graph-index, new model vectors, 1,000 queries, exact ground truth, and
out-degree bound $M=64$.  From each pair of serialized indexes, we
sample 1,000 fixed-seed live vertices; each graph contributes its
closest $\min(\deg_{\mathrm{Replace}}(u),\deg_{\mathrm{Local}}(u))$
bottom-layer edges per vertex.  We compute their edge-weighted mean new
model cosine distance, shown in  \Cref{fig:local_vs_replace}(a) after
normalization to Res-Replace result.  \Cref{fig:local_vs_replace}(b) shows the paired Recall@10
difference at $\mathrm{ef}=50$ and 100.

As \Cref{fig:local_vs_replace}(a) shows, Res-Local reduces the
degree-matched mean neighbor distance by 6.0\%--14.4\% relative to Res-Replace.
Its Recall@10
gain over Res-Replace is 3.36--22.66\,pp at
$\mathrm{ef}=50$ and 2.13--17.29\,pp at
$\mathrm{ef}=100$.  The gain is largest on Yelp, whose inherited
graph is least effective under the new model.

\begin{table*}[!t]
  \centering
  \caption{Res-Search Migration Time across Eight Workloads.}
  \label{tab:main_search_time}
  \ExpTableFont
  \begin{tabular*}{\textwidth}{@{\extracolsep{\fill}}ccccccc@{}}
    \toprule
    & \multicolumn{5}{c}{Migration time (s)} & Speedup range \\
    \cmidrule(lr){2-6}
    Dataset & Res-Search & HNSW & Vamana & NSG & NN-Descent & ($\times$) \\
    \midrule
    CIFAR-10    &    1.13 &    6.75 &    6.07 &   23.47 &   11.70 & 5.36--20.72 \\
    AG News     &   13.38 &   74.63 &   80.88 &  132.09 &   76.17 & 5.58--9.87 \\
    Yelp        &   54.25 &  208.47 &  241.80 &  523.38 &  486.41 & 3.84--9.65 \\
    Amazon-1536 &  186.11 &  455.05 &  422.27 & 1238.84 &  902.74 & 2.27--6.66 \\
    FineWeb     &   64.80 &  197.89 &  199.74 &  662.63 &  508.00 & 3.05--10.23 \\
    Yahoo       &  137.84 &  504.90 &  502.98 &  770.25 &  418.02 & 3.03--5.59 \\
    QuickDraw   &   60.00 &  257.64 &  265.68 &  653.34 &  440.53 & 4.29--10.89 \\
    MS\,MARCO   & 1045.53 & 3100.11 & 3280.64 & 7162.38 & 4391.88 & 2.97--6.85 \\
    \bottomrule
  \end{tabular*}
\end{table*}

\begin{figure*}[!t]
  \centering
  \includegraphics[width=\textwidth]{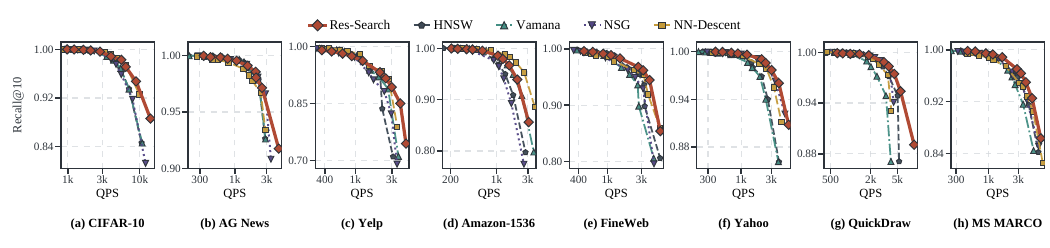}\vspace{-1ex}
  \caption{Res-Search Recall--QPS Trade-offs across Eight Workloads.}\vspace{-1ex}
  \label{fig:main_search_quality}
\end{figure*}

\subsection{Overall Performance of Res-Search}

\subsubsection{Refinement Quality on the Main Experimental Datasets}
\label{sec:exp:search_quality}

\Cref{tab:main_search_time,fig:main_search_quality} answer RQ2 by comparing
graph-processing time and measured Recall--QPS frontiers across the eight main
workloads.  Res-Search is
$2.27\times$--$5.58\times$ faster than the fastest reconstruction baseline and
$5.59\times$--$20.72\times$ faster than the slowest.  Across all workloads and search
settings, Res-Search improves averaged Recall@10 by 0.35 percentage points and
QPS by 22.3\% relative to the fastest reconstruction baseline.

\paragraph{Graph-reuse baseline.}
We also warm-start a conventional graph construction from the same valuable input:
each new-model adjacency list is initialized with $N_{G^\mathbf{o}}(u)$, then standard
NN-Descent runs for ten rounds using new model distances (sample
rate 0.2, $M=64$, 32 threads).  Across the matched sweep,
Res-Search achieves higher recall than warm-start NN-Descent at all
$8\times10$ $\mathrm{ef}$ values, with per-dataset maximum gains of
1.02--16.78\,pp; it is also $1.31\times$--$6.49\times$
faster on seven datasets and effectively tied on Yahoo; the median speedup is
$1.79\times$.  It therefore does not subsume ResGraph's
reachability-guided candidate selection.

\subsubsection{Does Residual Reachability Matter?}
\label{sec:exp:structure_control}

To validate the practical value of residual reachability, we randomly
permute the vertex IDs of the old index to construct an ID-permuted
graph that preserves its complete topology and degree distribution while
destroying the alignment between each object and its inherited
neighborhood; the aligned--permuted gap therefore isolates the value of
graph--vector alignment.  As an auxiliary control, a connected random
64-regular graph with no inherited topology tests whether the old
graph's unlabeled organization itself offers an advantage over a generic
scaffold for candidate propagation.  Starting from the aligned graph and
both controls, we run identical Res-Local and Res-Search pipelines, with
Search starting from the corresponding Res-Replace graph.
\Cref{tab:structure_value_controls} reports maximum Recall@10 loss for
Local and Search, together with the degree-matched mean cosine neighbor
distance $\bar d_N$ in the new embedding space at Res-Replace,
Res-Local, and Res-Search, measured on the same 1,000 sampled vertices
(lower is better).

\begin{table}[!t]
  \centering
  \caption{Residual-Reachability Controls.}
  \label{tab:structure_value_controls}
  \ExpTableFont
  \setlength{\tabcolsep}{1.8pt}
  \begin{tabular}{@{}ccccccc@{}}
    \toprule
    & & \multicolumn{3}{c}{Mean neighbor distance $\bar d_N$}
      & \multicolumn{2}{c}{Max. Recall@10 loss (pp)} \\
    \cmidrule(lr){3-5}\cmidrule(l){6-7}
    Dataset & Start & Replace & Local & Search & Local & Search \\
    \midrule
    \multirow[c]{3}{*}{QuickDraw}
      & Inherited   & 0.0796 & 0.0618 & 0.0497 &  0.00 &  0.00 \\
      & ID-permuted & 0.2128 & 0.1494 & 0.0680 & 88.84 & 39.09 \\
      & 64-regular  & 0.1816 & 0.1211 & 0.0769 & 86.53 & 59.26 \\
    \midrule
    \multirow[c]{3}{*}{MS\,MARCO}
      & Inherited   & 0.1420 & 0.1396 & 0.1243 &  0.00 &  0.00 \\
      & ID-permuted & 0.2825 & 0.2349 & 0.1723 & 96.04 & 51.41 \\
      & 64-regular  & 0.2699 & 0.2274 & 0.1893 & 96.13 & 78.58 \\
    \bottomrule
  \end{tabular}
\end{table}

Destroying ID alignment causes maximum Recall@10 losses of 86.53--96.13\,pp
for Res-Local and 39.09--78.58\,pp for Res-Search.  The neighbor-distance
measurement exposes the corresponding structural loss: relative to the
aligned inherited graph, the two controls increase $\bar d_N$ by
90.1\%--167.5\% at Replace and 62.9\%--141.6\% after Local.
After Search, their mean neighbor distances remain 36.7--54.7\% larger.
After refinement, shorter neighbor distances
for ID-permuted graphs than random regular graphs suggest that inherited graph
structure also contributes.
The aligned--permuted gap isolates graph--vector alignment, whose value is
further supported by the degraded 64-regular control.  ResGraph thus
benefits from residual reachability rather than merely from a convenient
initial graph.

\begin{figure*}[t]
  \begin{minipage}[t]{0.48\textwidth}
    \centering
    \includegraphics[width=\linewidth]{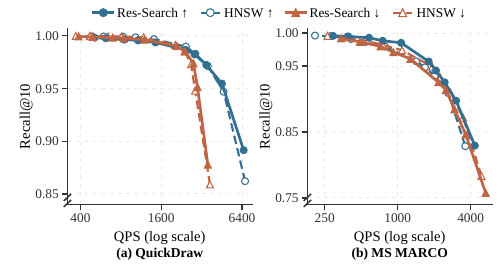}\vspace{-1ex}
    \captionof{figure}{Res-Search across Dimension Changes.}\vspace{-1ex}
    \label{fig:dimension_changing_migration}
  \end{minipage}\hfill
  \begin{minipage}[t]{0.48\textwidth}
    \centering
    \includegraphics[width=\linewidth,trim=2.5pt 2.5pt 2pt 2pt,clip]{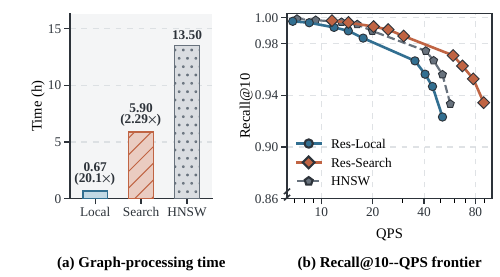}\vspace{-1ex}
    \captionof{figure}{ResGraph at 100 Million Vectors.}\vspace{-1ex}
    \label{fig:c4_100m_scale}
  \end{minipage}
\end{figure*}

\subsubsection{Scaling to 100 Million Vectors}
\label{sec:exp:c4_100m}

\Cref{fig:c4_100m_scale} tests ResGraph one order of magnitude beyond the largest
main workload.  Panel (a) compares graph-processing time with the HNSW
speedups; panel (b) shows the nondominated Recall@10--QPS points on a
truncated high-recall axis.

HNSW graph processing takes 13.501 hours, versus 0.673 for Res-Local and 5.898
for Res-Search, corresponding to speedups of $20.06\times$ and
$2.29\times$, respectively.
At $\mathrm{ef}=400$, their Recall@10 gaps to HNSW are only 0.18 and
0.13 percentage points.  At $\mathrm{ef}=50$, Res-Search is within 0.35
percentage points, while reducing latency by 30.8\% relative to
HNSW.  The graph-processing advantage and high endpoint quality therefore persist at
approximately 100M vectors.

\subsubsection{Generalization across Corpora and Model Pairs}

\Cref{fig:dimension_changing_migration} extends RQ3 beyond
$d_\mathbf{o}=d_\mathbf{n}$ with two image-embedding targets and two
passage-embedding targets.  Panel (a) overlays the QuickDraw
MetaCLIP-512-to-LAION-768 and MetaCLIP-512-to-DINOv2-384 replacements; panel
(b) overlays the MS~MARCO E5-768-to-Arctic-1024 and E5-768-to-MiniLM-384
replacements.  Each method--target curve contains the ten measured
$\mathrm{ef}$ settings.  Blue circles mark dimension increases and orange
triangles decreases; solid filled curves identify Res-Search, dashed hollow
ones HNSW.

Across these four replacements, Res-Search reduces graph-processing time by
$2.33\times$, $2.67\times$, $5.12\times$, and $3.19\times$, respectively.
At $\mathrm{ef}=400$, its Recall@10 is within 0.13, 0.05, 0.37, and 0.05
percentage points of HNSW, while its QPS is higher by 4.8\%, 5.0\%, 29.4\%,
and 40.8\%.  As the early-availability index, Res-Local is
$9.24\times$--$16.83\times$ faster than HNSW across the same four replacements and
reaches 97.70\%--99.78\% Recall@10 at $\mathrm{ef}=400$.  Thus ResGraph preserves
its availability/refinement trade-off across both dimension-increasing and
dimension-decreasing model replacements.
\ifarxivversion
Their complete results appear in
Appendix~\ref{app:generalization} and \Cref{fig:generalization}.
\fi
Together with
\Cref{tab:local_tradeoff,fig:c4_100m_scale,fig:dimension_changing_migration},
this answers RQ3: ResGraph's construction advantage spans all eight main workloads,
persists under dimension-changing replacements and at approximately 100M
vectors, and refinement helps every controlled comparison.

\subsection{Multi-Round Cross-Model Iteration}\label{sec:exp:iteration}

Deployed retrieval systems upgrade their embedding models repeatedly,
so the graph ResGraph migrates to model $\mathbb M_r$ becomes the legacy
input for the next upgrade to $\mathbb M_{r+1}$.  Two questions arise
that a single migration cannot answer: (i)~does residual reachability
persist in a graph whose edges were already re-selected for the
previous model, and (ii)~do migration errors accumulate over
successive rounds?

\paragraph{Chains and protocol.}
We instantiate the iterated migration of \Cref{eq:iterated_migration}
with ID-aligned released models on two corpora.  On Amazon, MPNet
$\rightarrow$ E5 $\rightarrow$ GTE $\rightarrow$ Qwen2-1.5B
$\rightarrow$ Qwen2.5-1.5B performs four upgrades over the same 999K
objects: two dimension-preserving encoder replacements, one
768-to-1536 change into decoder-based embeddings, and one
1536-dimensional round.  On Yahoo, E5 $\rightarrow$ GTE
$\rightarrow$ MultiQA-MPNet adds two 768-dimensional rounds on a second
corpus.  Round-0 indexes are HNSW graphs built on the oldest model under
the settings of \S\ref{sec:exp:setup}.  Every round applies the main
pipeline: Res-Replace, Res-Local from the chain graph with inline vector
replacement, and Res-Search from the Replace graph, whose output becomes
the chain graph of round $r+1$.  A fresh HNSW rebuild on the round-$r$
vectors is the per-round reference, and, from round two, a fresh-source
control reruns the round's ResGraph stages from the round-$(r{-}1)$ rebuild
instead of the chain graph, isolating the effect of an iterated source
graph.  Residual reachability per round is the exact top-10 hop coverage
of the Replace graph (\Cref{eq:hop_coverage}).

\begin{table}[!t]
  \centering
  \caption{Cross-Model Iteration across Two Model Chains.}
  \label{tab:iteration}
  \ExpTableFont
  \setlength{\tabcolsep}{2.6pt}
  \begin{tabular}{@{}llrrrrrr@{}}
    \toprule
    & & \multicolumn{3}{c}{Graph time (s)} & \multicolumn{2}{c}{$\Delta$R@10 (pp)} & $\mathrm{cov}_4$ \\
    \cmidrule(lr){3-5}\cmidrule(l){6-7}
    Round & Upgrade & Loc. & Srch. & HNSW & chain & fresh & (\%) \\
    \midrule
    A1 & MPNet$\rightarrow$E5 & 19.3 & 157.9 & 514.5 & 1.29 & -- & 98.8 \\
    A2 & E5$\rightarrow$GTE & 27.4 & 203.4 & 434.4 & 0.65 & 0.41 & 99.8 \\
    A3 & GTE$\rightarrow$Qwen2 & 38.8 & 137.7 & 338.3 & 1.39 & 0.68 & 99.2 \\
    A4 & Qwen2$\rightarrow$Qwen2.5 & 32.6 & 178.6 & 410.4 & 0.20 & 0.09 & 99.3 \\
    \addlinespace[0.15em]
    Y1 & E5$\rightarrow$GTE & 26.6 & 144.5 & 364.4 & 0.04 & -- & 100.0 \\
    Y2 & GTE$\rightarrow$MQA-MPNet & 29.6 & 113.4 & 394.4 & 0.40 & 0.20 & 100.0 \\
    \bottomrule
  \end{tabular}
\end{table}

\paragraph{Findings.}
\Cref{tab:iteration} supports three conclusions.  First, the
construction advantage persists in every round: Res-Search finishes
within 2.14$\times$--3.48$\times$ of the per-round rebuild time and
Res-Local within 8.73$\times$--26.66$\times$; summed over the four
Amazon upgrades, Res-Search spends 677.6\,s versus 1697.7\,s of
rebuilds.  Second, iteration does not erode residual reachability: the
Replace graphs retain $\mathrm{cov}_4\ge98.8\%$ in every round, including the dimension
change, and their two-hop coverage even exceeds the fresh-source
controls by up to 17.1\,pp.  Third, migration errors do not accumulate:
chain-sourced Res-Search is at most 1.39\,pp behind the same-round
rebuild over the whole sweep (0.82\,pp at $\mathrm{ef}=400$) and matches
the fresh-source control within 0.70\,pp (0.12\,pp at
$\mathrm{ef}=400$).  The hardest round is GTE$\rightarrow$Qwen2, whose
model pair shares only 6.1\% of exact top-10 neighbors: the
Replace-only graph collapses to 62.0\%/87.3\% Recall@10 at
$\mathrm{ef}=50$/$400$, whereas hop-bounded migration
restores 97.1\%/99.7\%, within 1.2/0.3\,pp of the rebuild;
avoiding migration instead leaves a 5.5--35.1\,pp Recall@10 gap at
$\mathrm{ef}=50$ in every round.

\subsection{Ablation Studies}

Having isolated adjacency migration in \S\ref{sec:exp:replace_control}, we
complete RQ4 by evaluating candidate screening and the hop ceiling.

\subsubsection{Effectiveness of Candidate Screening}
\label{sec:exp:screening_ablation}

\ifarxivversion
Detailed screening experiments appear in
Appendix~\ref{app:screening_ablation} and
\Cref{fig:screening_efficiency}.
\fi
At a matched top-400 shortlist budget, MPS retains
80.73\%--98.06\% of the exact top-256 candidates across AG News, Yelp, and Amazon-1536.
RaBitQ raises coverage but costs
$2.43\times$--$3.94\times$ more to score and requires training and full-dataset
encoding.  On MS\,MARCO, removing screening takes $2.15\times$ the
graph-processing time for Recall@10 gains of only 0.11 and 0.02 percentage
points at $\mathrm{ef}=50$ and 100.
MPS is therefore the primary temporary screen.

\subsubsection{Effect of Hop-Bounded Refinement}
\label{sec:exp:hop_ablation}

\begin{figure}[!t]
  \centering
  \includegraphics[width=\columnwidth]{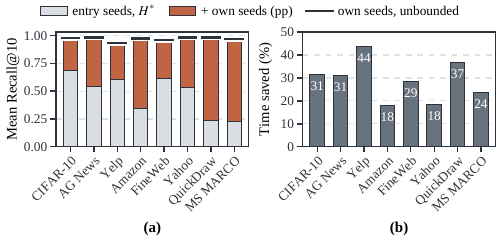}\vspace{-1ex}
  \caption{Seeding and Ceiling Contributions across Eight Workloads.}\vspace{-1ex}
  \label{fig:hop_ceiling_baseline}
\end{figure}

Refinement without a hop ceiling---every vertex is refined until its
beam frontier empties---is an explicit baseline on all eight workloads;
paired arms start from the same Res-Replace index and differ only in
depth admission.  The selected ceilings cut core refinement time by
17.9\%--43.7\% (median 29.9\%) while changing mean Recall@10 by
$-0.985$--$+0.240$\,pp over the ten shared $\mathrm{ef}$ values.  On
MS\,MARCO the
ceiling removes $9.7\times10^{9}$ exact distance evaluations and
$0.66\times10^{9}$ adjacency expansions.  Attributing the two
mechanisms requires re-seeding every beam from the global entry
vertex instead of the vertex's own neighbors:
\Cref{fig:hop_ceiling_baseline} reports both per workload---panel
(a) stacks the mean
Recall@10 of entry-seeded traversal under the selected ceiling (gray
base) with the additional recall that own-neighbor seeding
contributes (red top), the dash atop each column marking the
own-seeded baseline without a hop ceiling, while panel (b) reports
how much of that unbounded core-refinement time the selected ceiling
saves.  Without the ceiling,
entry seeding matches quality within $0.01$--$0.18$\,pp at
$1.03\times$--$1.17\times$ the refinement time once search is
unbounded; with the ceiling retained, entry seeding loses
$29.6$--$74.3$\,pp (absolute
$0.227$--$0.681$) while its truncated traversal terminates at only
$17$--$43\%$ of the time.  Own-neighbor seeding and the hop ceiling
are therefore complements: the seeds place each beam inside the region
that carries residual reachability, and the ceiling bounds the
remaining propagation at at most $1$\,pp.
Together, the three ablations answer RQ4: each mechanism reduces work while
preserving endpoint quality.

\subsection{Low-Budget Reconstruction}
\label{sec:exp:low_budget_rebuild}

\begin{figure}[!t]
  \centering
  \includegraphics[
    width=0.937\columnwidth,
    trim=11.93bp 7.72bp 3.14bp 13.75bp,
    clip
  ]{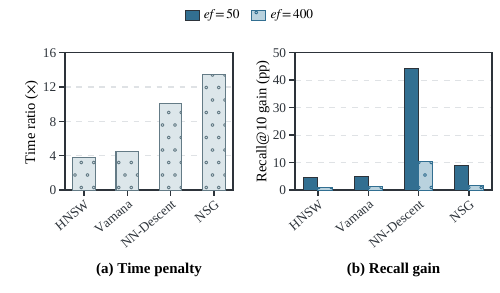}\vspace{-1ex}
  \caption{Low-Budget Reconstruction on MS\,MARCO.}\vspace{-1ex}
  \label{fig:low_budget_rebuild}
\end{figure}

MS\,MARCO (8.84M vectors) is our main workload closest to deployment scale.
We therefore stress-test unusually aggressive budgets: construction pool
$\mathrm{ef}_c=64$ for HNSW, width 64 for Vamana, five~NN-Descent
iterations, and an NSG initialization with five iterations followed by one
strict MRNG pass with $L=64$.
In \Cref{fig:low_budget_rebuild}, panel (a) reports each baseline's
graph-processing time relative to Res-Local, and panel (b) reports Res-Local's
Recall@10 gain at $\mathrm{ef}=50$ and 400.

Even at these budgets, the four reconstructions in
\Cref{fig:low_budget_rebuild} remain $3.77\times$--$13.44\times$ slower than
Res-Local and lower in Recall@10 by 4.78--44.31\,pp at $\mathrm{ef}=50$
(1.04--10.25\,pp at $\mathrm{ef}=400$).  Among these baselines,
NN-Descent exhibits the largest recall deficit, with Recall@10
44.31 percentage points below Res-Local at $\mathrm{ef}=50$ after five iterations.

Low-budget HNSW produces 696 connected components versus 240 for Res-Local,
and its largest component has 462 fewer vertices.  Its
mean exact top-10 adjacency recall---the fraction of exact top-10 corpus
neighbors present in each sampled neighbor list---is 4.17\%, compared with
18.12\% for Res-Local, a 77.0\% relative decline.  This fragmentation and
neighbor-quality loss quantify the cost of aggressive low-budget
construction: Res-Local pairs faster graph processing with stronger
new-space neighborhoods because the inherited graph supplies useful
candidate relationships for recovering them under limited budgets.

\section{Related Work}\label{sec:related}

Graph-based indexes offer leading search quality and efficiency for
high-dimensional approximate nearest neighbor search~\cite{wang2021survey}.
HNSW performs insertion-time graph
search, NN-Descent propagates candidates through neighbor joins, NSG and SSG
refine an intermediate neighbor graph, and Vamana combines graph search with
diversity-aware pruning
~\cite{malkov2020hnsw,dong2011nndescent,fu2019nsg,fu2022ssg,
subramanya2019diskann}.  Although the details differ, all discover and select
neighbors within a single embedding space; after corpus re-encoding, these
constructions must rediscover candidate relationships from the replacement
embeddings, leaving the relationships materialized in the old index unused.

Dynamic maintenance instead handles collection changes under a stable
embedding function.  HNSW supports online insertion, while SPFresh confines
updates and index reorganization to affected regions
~\cite{malkov2020hnsw,xu2023spfresh}.  Recent disk-resident designs likewise
localize graph repair or update adjacency lists in place
~\cite{yu2025topology,xu2025inplace}.  Insertions, deletions, and object
updates affect part of the collection without redefining existing
distances, whereas model replacement keeps object IDs fixed but moves every
vector and changes neighborhoods throughout the corpus, outside the scope
of localized update mechanisms.

Cross-model retrieval tackles the incompatibility between representations
produced by different model generations.  Representation-level methods
train a new encoder to remain compatible with old features, learn
transformations between embedding spaces, or describe samples relative
to shared anchors, with recent work further addressing successive
generations and query
drift~\cite{shen2020bct,hu2022compatible,moschella2023relative,yoon2025converter,
bui2025arrow,goswami2026query}.  At the database level, cross-model
integration supports retrieval over collections encoded by different
models, vector linking recovers correspondences between partially
overlapping embedding clouds, and Drift-Adapter maps new-model queries
into a legacy corpus space
~\cite{yang2025integrating,chen2026vectorlinking,vejendla2025drift}.  However,
these methods do not address how an existing index can be migrated to the
new space once the embedding model is replaced.

\makeatletter
\ifdefined\ACM@NRadjust
\let\DGMRegularSection\section
\renewcommand\section{\def\@toclevel{1}%
  \@startsection{section}{1}{\z@}%
  {-1.85\baselineskip \@minus -.2\p@}
  {.25\baselineskip}%
  {\ACM@NRadjust\@secfont}}
\fi
\section{Conclusion}\label{sec:conclusion}
\ifdefined\ACM@NRadjust
\let\section\DGMRegularSection
\fi
\makeatother

We study graph-index migration after embedding-model replacement, which
misaligns inherited edges with the new embedding space and makes full
reconstruction costly.  We observe residual reachability in the old graph
and propose ResGraph, which reuses the old graph for
candidate discovery and selects neighbors with exact new-space distances:
Res-Local combines parallel shallow expansion with packed-sign screening
for early availability, and Res-Search uses hop-bounded beam traversal to
recover candidates beyond local neighborhoods and approach
reconstruction-level quality.  Experiments on text and image migrations,
dimension-changing replacements, large-scale datasets, and repeated model
upgrades---where each migrated index becomes the next legacy
graph---demonstrate that ResGraph accelerates migration over full
reconstruction at competitive search quality, with its advantage and
residual reachability persisting over successive rounds without error
accumulation.

\FloatBarrier

\clearpage

\section*{Acknowledgment}
The authors disclose the use of OpenAI's ChatGPT to assist with drafting
the experimental code.  The experimental design,
testing, and correctness validation of all results were performed by the
authors.  ChatGPT was additionally used for editing and grammar polishing
of parts of the manuscript.  All AI-assisted content was reviewed and
verified by the authors, who take full responsibility for the
correctness and originality of this paper.


\bibliographystyle{IEEEtran}
\bibliography{references}

@article{malkov2020hnsw,
	author    = {Yury A. Malkov and Dmitry A. Yashunin},
	title     = {Efficient and Robust Approximate Nearest Neighbor Search Using
	Hierarchical Navigable Small World Graphs},
	journal   = {IEEE Transactions on Pattern Analysis and Machine Intelligence},
	volume    = {42},
	number    = {4},
	pages     = {824--836},
	year      = {2020},
	doi       = {10.1109/TPAMI.2018.2889473},
}

@inproceedings{dong2011nndescent,
	author    = {Wei Dong and Moses Charikar and Kai Li},
	title     = {Efficient $k$-Nearest Neighbor Graph Construction for Generic
	Similarity Measures},
	booktitle = {Proc.\ 20th Int.\ Conf.\ World Wide Web (WWW)},
	pages     = {577--586},
	publisher = {ACM},
	address   = {New York, NY, USA},
	year      = {2011},
	doi       = {10.1145/1963405.1963487},
}

@article{fu2019nsg,
	author    = {Cong Fu and Chao Xiang and Changxu Wang and Deng Cai},
	title     = {Fast Approximate Nearest Neighbor Search with the Navigating
	Spreading-out Graph},
	journal   = {Proceedings of the VLDB Endowment},
	volume    = {12},
	number    = {5},
	pages     = {461--474},
	year      = {2019},
	doi       = {10.14778/3303753.3303754},
}

@inproceedings{subramanya2019diskann,
	author    = {Suhas Jayaram Subramanya and {Devvrit} and
	Harsha Vardhan Simhadri and Ravishankar Krishnaswamy and
	Rohan Kadekodi},
	title     = {{DiskANN}: Fast Accurate Billion-point Nearest Neighbor
	Search on a Single Node},
	booktitle = {Advances in Neural Information Processing Systems},
	volume    = {32},
	pages     = {13748--13758},
	publisher = {Curran Associates, Inc.},
	address   = {Red Hook, NY, USA},
	year      = {2019},
}

@article{fu2022ssg,
	author    = {Cong Fu and Changxu Wang and Deng Cai},
	title     = {High Dimensional Similarity Search With Satellite System
	Graph: Efficiency, Scalability, and Unindexed Query Compatibility},
	journal   = {IEEE Transactions on Pattern Analysis and Machine Intelligence},
	volume    = {44},
	number    = {8},
	pages     = {4139--4150},
	year      = {2022},
	doi       = {10.1109/TPAMI.2021.3067706},
}

@article{wang2021survey,
	author    = {Mengzhao Wang and Xiaoliang Xu and Qiang Yue and Yuxiang Wang},
	title     = {A Comprehensive Survey and Experimental Comparison of
	Graph-Based Approximate Nearest Neighbor Search},
	journal   = {Proceedings of the VLDB Endowment},
	volume    = {14},
	number    = {11},
	pages     = {1964--1978},
	year      = {2021},
	doi       = {10.14778/3476249.3476255},
}

@article{zhong2025vsag,
	author    = {Xiaoyao Zhong and Haotian Li and Jiabao Jin and
	Mingyu Yang and Deming Chu and Xiangyu Wang and
	Zhitao Shen and Wei Jia and George Gu and Yi Xie and
	Xuemin Lin and Heng Tao Shen and Jingkuan Song and Peng Cheng},
	title     = {{VSAG}: An Optimized Search Framework for Graph-based
	Approximate Nearest Neighbor Search},
	journal   = {Proceedings of the VLDB Endowment},
	volume    = {18},
	number    = {12},
	pages     = {5017--5030},
	year      = {2025},
	doi       = {10.14778/3750601.3750624},
}

@article{gao2024rabitq,
	author    = {Jianyang Gao and Cheng Long},
	title     = {{RaBitQ}: Quantizing High-Dimensional Vectors with a
	Theoretical Error Bound for Approximate Nearest Neighbor Search},
	journal   = {Proceedings of the ACM on Management of Data},
	volume    = {2},
	number    = {3},
	articleno = {167},
	numpages  = {27},
	year      = {2024},
	doi       = {10.1145/3654970},
}

@inproceedings{xu2023spfresh,
	author    = {Yuming Xu and Hengyu Liang and Jin Li and Shuotao Xu and
	Qi Chen and Qianxi Zhang and Cheng Li and Ziyue Yang and
	Fan Yang and Yuqing Yang and Peng Cheng and Mao Yang},
	title     = {{SPFresh}: Incremental In-Place Update for Billion-Scale
	Vector Search},
	booktitle = {Proc. 29th Symp. Operating Systems Principles (SOSP)},
	pages     = {545--561},
	publisher = {ACM},
	address   = {New York, NY, USA},
	year      = {2023},
	doi       = {10.1145/3600006.3613166},
}

@article{yu2025topology,
	author    = {Song Yu and Shengyuan Lin and Shufeng Gong and Yongqing Xie and
	Ruicheng Liu and Yijie Zhou and Ji Sun and Yanfeng Zhang and
	Guoliang Li and Ge Yu},
	title     = {A Topology-Aware Localized Update Strategy for Graph-Based
	{ANN} Index},
	journal   = {Proceedings of the VLDB Endowment},
	volume    = {19},
	number    = {3},
	pages     = {495--508},
	year      = {2025},
	doi       = {10.14778/3778092.3778108},
}

@misc{xu2025inplace,
	author    = {Haike Xu and Magdalen Dobson Manohar and Philip A. Bernstein and
	Badrish Chandramouli and Richard Wen and Harsha Vardhan Simhadri},
	title     = {In-Place Updates of a Graph Index for Streaming Approximate
	Nearest Neighbor Search},
	howpublished = {arXiv preprint},
	eprint    = {2502.13826},
	archiveprefix = {arXiv},
	primaryclass = {cs.IR},
	year      = {2025},
	doi       = {10.48550/arXiv.2502.13826}
}

@inproceedings{karpukhin2020dense,
	author    = {Vladimir Karpukhin and Barlas O\u{g}uz and Sewon Min and
	Patrick Lewis and Ledell Wu and Sergey Edunov and
	Danqi Chen and Wen-tau Yih},
	title     = {Dense Passage Retrieval for Open-Domain Question Answering},
	booktitle = {Proc.\ Conf.\ Empirical Methods in Natural Language
	Processing (EMNLP)},
	pages     = {6769--6781},
	publisher = {Association for Computational Linguistics},
	address   = {Online},
	year      = {2020},
	doi       = {10.18653/v1/2020.emnlp-main.550},
}

@inproceedings{lewis2020rag,
	author    = {Patrick Lewis and Ethan Perez and Aleksandra Piktus and
	Fabio Petroni and Vladimir Karpukhin and Naman Goyal and
	Heinrich K\"{u}ttler and Mike Lewis and Wen-tau Yih and
	Tim Rockt\"{a}schel and Sebastian Riedel and Douwe Kiela},
	title     = {Retrieval-Augmented Generation for Knowledge-Intensive
	{NLP} Tasks},
	booktitle = {Advances in Neural Information Processing Systems},
	volume    = {33},
	pages     = {9459--9474},
	publisher = {Curran Associates, Inc.},
	address   = {Red Hook, NY, USA},
	year      = {2020},
}

@inproceedings{wang2020minilm,
	author    = {Wenhui Wang and Furu Wei and Li Dong and Hangbo Bao and
	Nan Yang and Ming Zhou},
	title     = {{MiniLM}: Deep Self-Attention Distillation for Task-Agnostic
	Compression of Pre-Trained Transformers},
	booktitle = {Advances in Neural Information Processing Systems},
	volume    = {33},
	pages     = {5776--5788},
	publisher = {Curran Associates, Inc.},
	address   = {Red Hook, NY, USA},
	year      = {2020},
}

@inproceedings{song2020mpnet,
	author    = {Kaitao Song and Xu Tan and Tao Qin and Jianfeng Lu and
	Tie-Yan Liu},
	title     = {{MPNet}: Masked and Permuted Pre-Training for Language
	Understanding},
	booktitle = {Advances in Neural Information Processing Systems},
	volume    = {33},
	pages     = {16857--16867},
	publisher = {Curran Associates, Inc.},
	address   = {Red Hook, NY, USA},
	year      = {2020},
}

@inproceedings{xiao2024cpack,
	author    = {Shitao Xiao and Zheng Liu and Peitian Zhang and
	Niklas Muennighoff and Defu Lian and Jian-Yun Nie},
	title     = {{C-Pack}: Packed Resources for General Chinese Embeddings},
	booktitle = {Proceedings of the 47th International ACM SIGIR Conference
	on Research and Development in Information Retrieval},
	pages     = {641--649},
	publisher = {ACM},
	address   = {New York, NY, USA},
	year      = {2024},
	doi       = {10.1145/3626772.3657878},
}

@inproceedings{lee2025nvembed,
	author    = {Chankyu Lee and Rajarshi Roy and Mengyao Xu and
	Jonathan Raiman and Mohammad Shoeybi and Bryan Catanzaro and Wei Ping},
	title     = {{NV-Embed}: Improved Techniques for Training {LLM}s as
	Generalist Embedding Models},
	booktitle = {The Thirteenth International Conference on Learning
	Representations},
	publisher = {OpenReview.net},
	address   = {Singapore},
	numpages  = {24},
	year      = {2025}
}

@techreport{krizhevsky2009cifar,
	author      = {Alex Krizhevsky and Geoffrey Hinton},
	title       = {Learning Multiple Layers of Features from Tiny Images},
	institution = {University of Toronto},
	year        = {2009},
	url         = {https://www.cs.toronto.edu/~kriz/learning-features-2009-TR.pdf}
}

@inproceedings{zhang2015charcnn,
	author    = {Xiang Zhang and Junbo Zhao and Yann LeCun},
	title     = {Character-level Convolutional Networks for Text
	Classification},
	booktitle = {Advances in Neural Information Processing Systems},
	volume    = {28},
	pages     = {649--657},
	publisher = {Curran Associates, Inc.},
	address   = {Red Hook, NY, USA},
	year      = {2015},
}

@inproceedings{penedo2024fineweb,
	author    = {Guilherme Penedo and Hynek Kydl{\'i}{\v{c}}ek and
	Loubna Ben Allal and Anton Lozhkov and Margaret Mitchell and
	Colin Raffel and Leandro von Werra and Thomas Wolf},
	title     = {The {FineWeb} Datasets: Decanting the Web for the Finest Text
	Data at Scale},
	booktitle = {Advances in Neural Information Processing Systems},
	volume    = {37},
	pages     = {30811--30849},
	publisher = {Curran Associates, Inc.},
	address   = {Red Hook, NY, USA},
	year      = {2024},
	doi       = {10.52202/079017-0970},
}

@article{raffel2020t5,
	author  = {Colin Raffel and Noam Shazeer and Adam Roberts and Katherine Lee
	and Sharan Narang and Michael Matena and Yanqi Zhou and Wei Li and
	Peter J. Liu},
	title   = {Exploring the Limits of Transfer Learning with a Unified
	Text-to-Text Transformer},
	journal = {Journal of Machine Learning Research},
	volume  = {21},
	number  = {140},
	pages   = {1--67},
	year    = {2020}
}

@inproceedings{nguyen2016msmarco,
	author    = {Tri Nguyen and Mir Rosenberg and Xia Song and Jianfeng Gao and
	Saurabh Tiwary and Rangan Majumder and Li Deng},
	title     = {{MS MARCO}: A Human-Generated {MA}chine Reading
	{CO}mprehension
	Dataset},
	booktitle = {Proceedings of the Workshop on Cognitive Computation:
	Integrating Neural and Symbolic Approaches 2016},
	series    = {CEUR Workshop Proceedings},
	volume    = {1773},
	publisher = {CEUR-WS.org},
	address   = {Barcelona, Spain},
	numpages  = {10},
	year      = {2016}
}

@inproceedings{he2016resnet,
	author    = {Kaiming He and Xiangyu Zhang and Shaoqing Ren and Jian Sun},
	title     = {Deep Residual Learning for Image Recognition},
	booktitle = {Proceedings of the IEEE Conference on Computer Vision and
	Pattern Recognition (CVPR)},
	pages     = {770--778},
	publisher = {IEEE Computer Society},
	address   = {Las Vegas, NV, USA},
	year      = {2016},
	doi       = {10.1109/CVPR.2016.90},
}

@inproceedings{covington2016deep,
	author    = {Paul Covington and Jay Adams and Emre Sargin},
	title     = {Deep Neural Networks for {YouTube} Recommendations},
	booktitle = {Proc.\ 10th ACM Conf.\ Recommender Systems (RecSys)},
	pages     = {191--198},
	publisher = {ACM},
	address   = {New York, NY, USA},
	year      = {2016},
	doi       = {10.1145/2959100.2959190},
}

@inproceedings{moschella2023relative,
	author    = {Luca Moschella and Valentino Maiorca and Marco Fumero and
	Antonio Norelli and Francesco Locatello and
	Emanuele Rodol\`{a}},
	title     = {Relative Representations Enable Zero-Shot Latent Space
	Communication},
	booktitle = {The Eleventh International Conference on Learning
	Representations},
	publisher = {OpenReview.net},
	address   = {Kigali, Rwanda},
	numpages  = {26},
	year      = {2023}
}

@inproceedings{shen2020bct,
	author    = {Yantao Shen and Yuanjun Xiong and Wei Xia and
	Stefano Soatto},
	title     = {Towards Backward-Compatible Representation Learning},
	booktitle = {Proc.\ IEEE/CVF Conf.\ Computer Vision and Pattern
	Recognition (CVPR)},
	pages     = {6368--6377},
	publisher = {Computer Vision Foundation / IEEE},
	address   = {Seattle, WA, USA},
	year      = {2020},
	doi       = {10.1109/CVPR42600.2020.00640},
}

@inproceedings{hu2022compatible,
	author    = {Weihua Hu and Rajas Bansal and Kaidi Cao and Nikhil Rao and
	Karthik Subbian and Jure Leskovec},
	title     = {Learning Backward Compatible Embeddings},
	booktitle = {Proc. 28th ACM SIGKDD Conf. Knowledge Discovery and Data
	Mining (KDD)},
	pages     = {3018--3028},
	publisher = {ACM},
	address   = {New York, NY, USA},
	year      = {2022},
	doi       = {10.1145/3534678.3539194},
}

@inproceedings{yoon2025converter,
	author    = {Jinsung Yoon and Sercan O. Arik},
	title     = {Embedding-Converter: A Unified Framework for Cross-Model
	Embedding Transformation},
	booktitle = {Proceedings of the 63rd Annual Meeting of the Association for
	Computational Linguistics (Volume 1: Long Papers)},
	pages     = {25464--25482},
	publisher = {Association for Computational Linguistics},
	address   = {Vienna, Austria},
	year      = {2025}
}

@inproceedings{bui2025arrow,
	author    = {Ngoc Bui and Menglin Yang and Runjin Chen and Leonardo Neves and
	Mingxuan Ju and Zhitao Ying and Neil Shah and Tong Zhao},
	title     = {Learning Along the Arrow of Time: Hyperbolic Geometry for
	Backward-Compatible Representation Learning},
	booktitle = {Proceedings of the 42nd International Conference on Machine
	Learning},
	series    = {Proceedings of Machine Learning Research},
	volume    = {267},
	pages     = {5842--5855},
	publisher = {PMLR},
	year      = {2025}
}

@inproceedings{goswami2026query,
	author    = {Dipam Goswami and Liying Wang and Bart{\l}omiej Twardowski and
	Joost van de Weijer},
	title     = {Query Drift Compensation: Enabling Compatibility in Continual
	Learning of Retrieval Embedding Models},
	booktitle = {Proceedings of The 4th Conference on Lifelong Learning Agents},
	series    = {Proceedings of Machine Learning Research},
	volume    = {330},
	pages     = {324--340},
	publisher = {PMLR},
	year      = {2026}
}

@inproceedings{vejendla2025drift,
	author    = {Harshil Vejendla},
	title     = {Drift-Adapter: A Practical Approach to Near Zero-Downtime
	Embedding Model Upgrades in Vector Databases},
	booktitle = {Proc. Conf. Empirical Methods in Natural Language Processing
	(EMNLP)},
	pages     = {15938--15949},
	publisher = {Association for Computational Linguistics},
	address   = {Suzhou, China},
	year      = {2025},
	doi       = {10.18653/v1/2025.emnlp-main.805},
}

@article{yang2025integrating,
	author    = {Beining Yang and Yang Cao and Yang Ren},
	title     = {Integrating Vector Databases across Embedding Models},
	journal   = {Proceedings of the ACM on Management of Data},
	volume    = {3},
	number    = {6},
	articleno = {338},
	numpages  = {28},
	year      = {2025},
	doi       = {10.1145/3769803},
}

@inproceedings{
	chen2026vectorlinking,
	title={Vector Linking via Cross-Model Local Isometric Consistency},
	author={Ziying Chen and Yang Cao and He Sun and Beining Yang and Tianjian Yang},
	booktitle={Forty-third International Conference on Machine Learning},
	year={2026},
	url={https://openreview.net/forum?id=wSV1sGTCN0}
}

@misc{yang2024qwen2,
	author    = {An Yang and others},
	title     = {{Qwen2} Technical Report},
	howpublished = {arXiv preprint},
	numpages  = {26},
	eprint    = {2407.10671},
	archiveprefix = {arXiv},
	primaryclass = {cs.CL},
	doi       = {10.48550/arXiv.2407.10671},
	year      = {2024},
}

@misc{qwen2024qwen25,
	author    = {An Yang and others},
	title     = {{Qwen2.5} Technical Report},
	howpublished = {arXiv preprint},
	numpages  = {26},
	eprint    = {2412.15115},
	archiveprefix = {arXiv},
	primaryclass = {cs.CL},
	doi       = {10.48550/arXiv.2412.15115},
	year      = {2024},
}

@inproceedings{ha2018neural,
	author    = {David Ha and Douglas Eck},
	title     = {A Neural Representation of Sketch Drawings},
	booktitle = {International Conference on Learning Representations (ICLR)},
	year      = {2018},
}

@inproceedings{radford2021clip,
	author    = {Alec Radford and others},
	title     = {Learning Transferable Visual Models from Natural Language
	Supervision},
	booktitle = {Proceedings of the 38th International Conference on Machine
	Learning (ICML)},
	pages     = {8748--8763},
	year      = {2021},
}

\end{document}